\documentclass[11pt]{article}
\usepackage{amsmath,amssymb}
\usepackage{hyperref}
\usepackage{rotating}
\usepackage{xypic}
\usepackage{stmaryrd}
\usepackage{tikz}
\usepackage{tkz-graph}
\usepackage{tkz-berge}
\usepackage{amssymb}
\usepackage{amsmath,amscd}
\usepackage{pst-node,pstricks,multido,pst-plot,pst-text,pst-3d}%
\usepackage{graphicx}
\usepackage{amsfonts}
\newtheorem{example}{Example}[section]

\newtheorem{remark}[example]{Remark}
\newtheorem{theorem}[example]{Theorem}

\newtheorem{corollary}[example]{Corollary}

\newtheorem{proposition}[example]{Proposition}

\newtheorem{lemma}[example]{Lemma}

\def\qed{\hspace{3.5mm} \hfill $\Box$
\bigskip\\}

\def\S{{\mathfrak  S}}

\def\<{\langle}
\def\>{\rangle}

\def\N{{\mathbb N}}

\newcommand{\spec}{\text{Spec}}

\def\ashuff#1#2#3{
\kern 1pt \vrule height#1 \overline{\vrule height#3 width 0pt
\hskip#2} \rule{.3pt}{#1}\overline{\vrule height#3 width 0pt
\hskip#2} \rule{.3pt}{#1} \kern 1pt }

\def\Carre3#1{\left[\begin{array}{ccc}#1\end{array}\right]}

\include {mak}
\begin{document}

\title{Factorizations of symmetric Macdonald polynomials}
\author{Laura Colmenarejo\footnote{Laura Colmenarejo, 
Department of Mathematics and Statistics, York University , Toronto, Canada. email: {\tt laura.colmenarejo.hernando@gmail.com}}, 
Charles F. Dunkl\footnote{Charles F. Dunkl, Department of Mathematics, University of Virginia,
Charlottesville VA 22904-4137, USA.
home page: {\tt http://people.virginia.edu/$\sim$cfd5z/},
e-mail: {\tt cfd5z@virginia.edu}}, and Jean-Gabriel Luque\footnote{Jean-Gabriel Luque, Universit\'e de Rouen Normandie,
Laboratoire d’Informatique, du Traitement de l’Information et des Syst\`emes (LITIS),
Avenue de l'Universit\'e - BP 8, 76801 Saint-\'Etienne-du-Rouvray Cedex, France. email:{\tt jean-gabriel.luque@univ-rouen.fr}}}
\maketitle{}
\hfill{\it D\'edicac\'e \`a Jean-Yves Thibon pour son soixanti\`eme anniversaire}
\begin{abstract}
	We prove many factorization formulas for highest weight Macdonald polynomials indexed by particular partitions called
	quasistaircases. As a consequence we prove a conjecture of Bernevig and Haldane stated in the context of the fractional quantum Hall theory.
\end{abstract}
\section{Introduction}
Jack polynomials have many applications in physics, in particular in statistical physics and quantum physics, due to their relation to the many-body problem. In particular, fractional quantum Hall states
of particles in the lowest Landau levels are described by such polynomials \cite{BH1,BH2,BH3}. In that context, some properties, called clustering
properties, are highly relevant. A clustering property can  occur for some negative rational parameters of a Jack polynomial and 
means that the 
Jack polynomial vanishes when $s$ distinct clusters of $k+1$ equal variables are formed. Using tools of algebraic geometry,
 Berkesch-Zamaere et al proved several
clustering properties \cite{BGS} including some special cases conjectured by Bervenig and Haldane \cite{BH1}. Coming from theoretical physics, the study of these 
properties raises very interesting problems in combinatorics and representation theory of the affine Hecke algebras.
 More precisely, the problem
is studied in the realm of Macdonald polynomials which form a $(q,t)$-deformation of the Jack polynomials 
related to the double affine Hecke algebras and the results are recovered by making degenerate the parameters $q$ and $t$.  
Instead of stating the results in terms of clustering properties, we prefer to state them in terms of factorizations. Indeed, clustering properties
are shown to be equivalent to very elegant formulas involving factorizations of Macdonald polynomials. For instance, many such factorizations have been already investigated in  \cite{BBL,BL,BaraF,DL3}. 
In particular, this paper is the sequel of \cite{DL3} in which two of the authors prove factorizations for rectangular Macdonald polynomials. In this paper, we investigate factorizations for more general partitions, called quasistaircases. 

The paper is organized as follows.  In section \ref{Background} we recall essential prerequisite on Macdonald polynomials. In section \ref{Hall} we give a brief account on the physics motivations coming from the fractional Quantum Hall theory. In Section \ref{FactGen} we investigate some factorizations involved for generic values of $(q,t)$.
 Section \ref{Spec} is devoted to the special cases of specializations of the type $t^{\alpha}q^{\beta}=1$ and, in particular,
  to the consequences on spectral vectors. In Section \ref{wheel}, we deduce factorizations from the results of Feigen et al \cite{FM} and, in Section \ref{notwheel},
 we prove more general results which are consequences of the highest weight condition of some quasistaircase Macdonald polynomials proved in \cite{JL}.
 In the last section (Section \ref{Concl}), we first illustrate our results by proving a conjecture stated by Bernevig and Haldane \cite{BH2}
 and also we show many other examples of factorizations that do not follow from our formulas.
\section{Background\label{Background}}
This paper is focused in the study of four variants of the Macdonald polynomials: symmetric, nonsymmetric, shifted symmetric and shifted nonsymmetric. Before getting into the results, we introduce these polynomials as well as some useful notation.
All the results contained in this section are well known results showed in several papers (see eg \cite{BF1,BF2,Chered,Dunkl,KN,Knop1,Knop2,Okoun1,Sahi1}).
The results of \cite{Lascoux01,Lascoux03,Lascoux08} are extensively used throughout our paper.
\subsection{Partitions and Vectors}
A \emph{partition} $\lambda =(\lambda_1,\dots,\lambda_\ell)$ of $n$ is a weakly decreasing sequence of positive integers such that $\sum_{i} \lambda_i=n$. 
The length of a partition $\lambda$ is $\ell(\lambda)=\max\{i: \lambda_i>0\}$. 

We consider a \emph{dominance order} on the partitions: 
Let $\lambda$ and $\mu$ be partitions, we say that 
\begin{displaymath}
\lambda \preceq_D \mu \hspace{0.3cm} \text{ if and only if } \hspace{0.3cm}  \forall i,\ \lambda_1+\dots+\lambda_i \leq \mu_1+\dots + \mu_i.
\end{displaymath}

If we consider vectors instead of partitions, the notation is as follows: $v=\left[v[1],\dots,v[N]\right]$ is a vector of length $N$. Note that for vectors, the zero parts are taken into account in the length of the vector. We denote by $v^+$ the unique non increasing partition which is a permutation of $v$. We can consider the following \emph{standardization of $v$}, $\text{std}_v$: we label with integer from 0 to $N-1$ the positions in $v$ from the smallest entries to the largest one and from right to left. We define the \emph{reciprocal vector} of $v$ as $\langle v \rangle_{q,t} = \left[t^{\text{std}_v[1]}q^{v[1]}, \dots, t^{\text{std}_v[N]}q^{v[N]}\right]$, and the \emph{reciprocal sum} $ \displaystyle{\lbag v\rbag_{q,t}=\sum_{i=1}^{N}\langle v \rangle_{q,t}[i]}$. If there is no ambiguity, the indices $q$ and $t$ are omitted.

For example, consider the vector $v=[1,2,2,0,1,1]$ of length 6. Then, $\text{std}_v= [3,5,4,0,2,1]$, $\langle v \rangle = \left[ qt^3,q^2t^5,q^2t^4,1,qt^2,qt\right]$, and $ \displaystyle{\lbag v\rbag_{q,t}= 1+qt+qt^2+qt^3+q^2t^4+q^2t^5}$.

Moreover, if $v=\lambda$ is a partition, $ \displaystyle{\lbag\lambda\rbag_{q,t}=\sum_{i=1}^{N}t^{N-i}q^{\lambda_{i}}=\sum_{i}\langle\lambda\rangle_{q,t}[i]}$.

The dominance order defined on the partitions is naturally extended to vectors almost with the same definition:
\begin{displaymath}
u \preceq v \hspace{0.3cm} \text{ if and only if  either } \hspace{0.3cm} u^+ \preceq_D v^+ \text{ or } \left( u^+=v^+ \text{ and } u \preceq_D v\right). 
\end{displaymath}

Note that this dominance order is defined initially only for vectors with the same norm. We can straightforwardly extend it for any vectors by adding the condition $u \prec v$ when $|u|<|v|$.

\subsection{Affine Hecke Algebra}

Let $N\geq 2$ be an integer, $t$ and $q$ be two independent parameters, and $\mathbb{X}=\{x_1,\dots,x_N\}$ be an alphabet of formal variables. We consider the right operators $T_i$ acting on Laurent polynomials in the variables $x_j$ by
\begin{eqnarray}\label{TiDef}
T_i=t+(s_i-1)\frac{tx_{i+1}-x_i}{x_{i+1}-x_i},
\end{eqnarray}
where $s_i$ is the elementary transposition permuting the variables $x_i$ and $x_{i+1}$. For instance, 
\begin{eqnarray}\label{TiEx}
1T_i = t \hspace{0.5cm} \text{and} \hspace{0.5cm} x_{i+1}T_i = x_i.
\end{eqnarray}

In fact, the operators $T_i$ are the unique operators that commute with multiplication  by symmetric functions in $x_i$ and $x_{i+1}$ satisfying \eqref{TiEx}. 

Consider also the affine operator $\tau$ defined by 
\begin{displaymath}
f(x_1,\dots,x_N)\tau = f\left(\frac{x_N}{q},x_1,\dots,x_{N-1}\right).
\end{displaymath}

The operators $T_i$ satisfy the relations of the Hecke algebra of the symmetric group:
\begin{eqnarray*}
& &T_iT_{i+1}T_i = T_{i+1}T_iT_{i+1}, \\
& &T_iT_j =T_jT_i \text{ for } |i-j|>1, \\
& &(T_i-t)(T_i+1)=0.
\end{eqnarray*}
Then, together with the multiplication by variables $x_i$ and the affine operator $\tau$, they generate the affine Hecke algebra of the symmetric group:
\begin{displaymath}
\mathcal{H}_N(q,t) = \mathbb{C}(q,t)\left[x_1^{\pm},\dots,x_N^{\pm},T_1^\pm,\dots, T_{N_1}^\pm, \tau\right].
\end{displaymath}

\subsection{Symmetric functions and virtual alphabets}
For the sake of simplicity, we will use $\Lambda$-ring notation for specializations of symmetric functions, see \cite{Lascoux03}.
By specialization we mean a morphism of algebra from $Sym$ to a commutative algebra. Since we manipulate only finite alphabets, specializing an alphabet is
equivalent to send each letters to a value. Notice that this is no longer the case for infinite alphabets for which the theory is more complicated.

 More precisely, we adopt the following
convention stated in terms of power sums. For any variable $x$, alphabets  $\mathbb X$ and $\mathbb Y$ and scalar $\alpha$, we set
\[p_{n}(x)=x^{n},\]
\[{}
p_{n}(\mathbb X+\mathbb Y)=p_{n}(\mathbb X)+p_{n}(\mathbb Y),
\]
\[{}
p_{n}(\mathbb X-\mathbb Y)=p_{n}(\mathbb X)-p_{n}(\mathbb Y),
\]
\[{}
p_{n}(\mathbb X\mathbb Y)=p_{n}(\mathbb X)p_{n}(\mathbb Y),
\]
\[{}
p_{n}(\alpha\mathbb Y)=\alpha \cdot  p_{n}(\mathbb Y).
\]
With this notation $\displaystyle{1-q^{n}\over 1-q}$ corresponds to the alphabet $\{1,q,\dots,q^{n-1}\}$. 
We set also $\mathbb X_{k}=\{x_{1},\dots,x_{k}\}$ and $\mathbb Y_{a,b}=\{y_{a},y_{a+1},\dots,y_{b}\}$, for $a\leq b$.\\
If $\mathbb X$ and $\mathbb Y$ are two alphabets, we will denote by 
$$\mathcal R(\mathbb X;\mathbb Y)=\prod_{(x,y)\in \mathbb X\times\mathbb Y}(x-y)$$ the resultant of $\mathbb X$ and $\mathbb Y$.
Since this is a symmetric function in  $\mathbb X$ and $\mathbb Y$ separately (but not in $\mathbb X\cup \mathbb Y$), we can use the notation above.
Hence one has
\[{}
\mathcal R(\mathbb X+\mathbb X';\mathbb Y)=\mathcal R(\mathbb X;\mathbb Y)\mathcal R(\mathbb X';\mathbb Y) \hspace{0.3cm}\mbox{ and } \hspace{0.3cm}
\mathcal R(\mathbb X;\mathbb Y+\mathbb Y')=\mathcal R(\mathbb X;\mathbb Y)\mathcal R(\mathbb X;\mathbb Y').
\]

\subsection{Macdonald Polynomials and Variants}

In this section, we set up the definitions and the notation for the Macdonald polynomials for the different variants that appear in the paper. We also define these variants for the Jack polynomials. 

Throughout this paper, the following notation is relevant and very useful. Let $P(\mathbb X; q,t)$ and $Q(\mathbb X; q,t)$ be two polynomials. We say that $P(\mathbb X; q,t)\displaystyle\mathop=^{(*)}Q(\mathbb X; q,t)$ if the equality holds up to a scalar factor consisting of powers of $q$ and $t$.  

\subsubsection*{Non symmetric (shifted) Macdonald polynomials}
The \emph{$(q,t)$-version of the Cherednik operators} are the operators defined by 
\begin{displaymath}
\xi_i:= t^{1-i}T_{i-1}\dots T_1 \tau T_{N-1}^{-1}\dots T_i^{-1}.
\end{displaymath}

The \emph{non symmetric Macdonald polynomials} $\left(E_v\right)_{v\in\mathbb{N}^N}$ are the unique basis of simultaneous eigenfunctions of the $(q,t)$-version of the Cherednik operators such that $E_v\displaystyle\mathop=^{(*)} x^v +\sum_{u \prec v} \alpha_{u,v}x^u$. The corresponding spectral vectors are given by the spectral vector, $\spec_v=\left(\frac{1}{\langle v \rangle [i]}\right)_{i=1}^N$. 

We consider also the following variant of the $\xi_i$ operators, the \emph{Knop-Cherednik operators}:
\begin{displaymath}
\Xi_i:= t^{1-i}T_{i-1}\dots T_1 \tau \left(1-\frac{1}{x_N}\right) T_{N-1}^{-1}\dots T_i^{-1} + \frac{1}{x_i}.
\end{displaymath}

The \emph{non symmetric shifted Macdonald polynomials} $\left(M_v\right)_{v\in\mathbb{N}^N}$ are the unique basis of simultaneous eigenfunctions of the Knop-Cherednik operators such that $M_v\displaystyle\mathop=^{(*)} x^v +\sum_{u \prec v} \alpha_{u,v}x^u$. As in the case of the non symmetric Macdonald polynomials, the spectral vector equals $\spec_v$.

Note that the polynomial $E_v$ can be recovered as a limit  from $M_v$:
\begin{displaymath}
E_v(x_1,\dots,x_N)= \lim_{a\rightarrow 0} a^{|v|}M_v\left(\frac{x_1}{a},\dots,\frac{x_N}{a}\right).
\end{displaymath}
This follows from the following fact:
\begin{equation*}
M_v(x_1,\dots,x_N)\mathop=^{(*)}E_v(x_1,\dots,x_N)+\sum_{u\prec v}\beta_{u,v}E_{u}.
\end{equation*}
The differences of the Cherednik operators and the Knop-Cherednik operators are known as the \emph{Dunkl operators}, $D_i = \Xi_i - \xi_i$. We say that a polynomial is \emph{singular} if it is in the kernel of $D_i$, for each $i$. 

\subsubsection*{Symmetric (shifted) Macdonald polynomials}
Let's consider the \emph{Debiard-Sekiguchi-Macdonald operator}, $\xi= \sum_i \xi_i$.

Then, the \emph{symmetric Macdonald polynomials}, $P_\lambda$, are defined as the eigenfunctions of $\xi$.

Similarly, we can consider the operator $\Xi=\sum_i \Xi_i$. Then, 
 the \emph{symmetric shifted Macdonald polynomials}, $MS_\lambda$, are defined as the eigenfunctions of $\Xi$.

The eigenvalue corresponding to the partition $\lambda$ is, in both cases,  $\lbag\lambda\rbag_{q^{-1},t^{-1}}$. 
We say that a polynomial satisfy the \emph{highest weight condition} if it is in the kernel of $\Xi-\xi$.
\subsubsection*{Jack polynomials}

We define the Jack polynomials, $J_v^\alpha$, as a limit of the Macdonald polynomials, $P_v$, with $q=t^\alpha$ and $t\rightarrow 1$. This definition applies for the four versions of Macdonald polynomials that appear in this paper.

\subsection{The Yang-Baxter graph}

In \cite{Lascoux01}, A. Lascoux described how to compute the non symmetric shifted Macdonald polynomials $M_v$ using the Yang-Baxter graph.
 This computation is based on the following result.
\begin{proposition}
\begin{itemize}
\item If $v[i]<v[i+1]$, 
\[
M_{v.s_i} = M_v \left( T_i + \frac{1-t}{1-\frac{\langle v \rangle[i+1]}{\langle v \rangle [i]}} \right),
\]
where $v.s_i$ is the vector obtained from $v$ by exchanging the values $v[i]$ and $v[i+1]$. 
\item $M_{v\Phi} = M_v \tau (x_N-1)$, where $v\Phi = \left[v[2],\dots,v[N],v[1]+1\right]$. 
\end{itemize}
\end{proposition}
This result provides a method to compute the polynomials $M_v$ following the Yang-Baxter graph associated to the vector $v$, 
starting with the zero vector $\left[ 0^N\right]$ and $M_{0^{N}}=1$. We illustrate how to do it for the non symmetric shifted Macdonald polynomials with an example. On one side, we have the following sequence for the vectors:
\begin{eqnarray*}
[000]  \xrightarrow{\Phi}   [001]  \xrightarrow{(23)}   [010]  \xrightarrow{(12)}  [100]  \xrightarrow{\Phi}   [002]  \xrightarrow{(23)}  [020]  \xrightarrow{\Phi}   [201].
\end{eqnarray*} 
It corresponds to the following sequence in the non symmetric shifted polynomials:
\begin{multline*}
M_{[000]}  \xrightarrow{\tau(x_3-1)}   M_{[001]}  \xrightarrow{T_2 + \frac{1-t}{1-qt^2}}   M_{[010]}  \xrightarrow{T_1 + \frac{1-t}{1-qt}}  M_{[100]}  \xrightarrow{\tau(x_3-1)} \\ \rightarrow M_{[002]} \xrightarrow{T_2 + \frac{1-t}{1-q^2t^2}}    M_{[020]}  \xrightarrow{\tau(x_3-1)}   M_{[201]}.
\end{multline*}
%The final result is
%\[ 
%M_{201}= 
%\]
The non symmetric (non shifted) Macdonald polynomials are obtained following an almost similar algorithm where the affine action is substituted
by $E_{v\Phi} = E_v \tau x_N$.
For instance,
\begin{multline*}
E_{[000]}  \xrightarrow{\tau x_3}   E_{[001]}  \xrightarrow{T_2 + \frac{1-t}{1-qt^2}}   E_{[010]}  
\xrightarrow{T_1 + \frac{1-t}{1-qt}}  E_{[100]}  \xrightarrow{\tau x_3}  E_{[002]} \xrightarrow{T_2 + \frac{1-t}{1-q^2t^2}}
    E_{[020]}  \xrightarrow{\tau x_3}   E_{[201]}.
\end{multline*}

The symmetric (shifted and non-shifted) are hence obtained by applying the  symmetrizing operator 
$\mathcal S=\sum_{\sigma\in\S _{N}}T_{\sigma}$, where $T_{\sigma}=T_{i_{1}}\cdots T_{i_{k}}$ if $\sigma=s_{i_{1}}\cdots s_{i_{k}}$ is a shortest
decomposition of $\sigma$ in elementary transposition $s_{i}=(i,\ i+1)$, to
the polynomials labeled by a decreasing vector.
Also Jack polynomials are  obtained following a Yang-Baxter graph with degenerated intertwining operators. 

\subsection{Vanishing properties}
The shifted polynomials in all their versions (non symmetric Macdonald, symmetric Macdonald, non symmetric Jack, symmetric Jack) can be defined alternatively by interpolation.
Indeed, one shows with the help of the Yang-Baxter graph that the shifted non symmetric Macdonald polynomials
are characterized up to a global coefficient by the equations
\begin{equation}\label{vanM}
M_{v}(\langle u\rangle)=0,
\end{equation}
for any vector $u$ satisfying $|u|\leq |v|$  and $u\neq v$.
By symmetrization, one shows that the shifted symmetric Macdonald polynomials are characterized up to a global coefficient 
by
\begin{equation}\label{vanSM}
MS_{\mu}(\langle \lambda\rangle)=0 \hspace{0.3cm} \mbox{ for } \hspace{0.3cm} |\lambda|\leq |\mu|\mbox{ and }\lambda\neq \mu,
\end{equation}
for any decreasing partition $\mu$.\\% satisfying $|\mu|\leq |\lambda|$  and $\mu\neq \lambda$.\\
Also, vanishing properties of shifted symmetric and non symmetric Jack polynomials are obtained by making equations (\ref{vanM}) and
(\ref{vanSM}) degenerate.

\section{Clustering properties of Jack polynomial and the quantum Hall effect \label{Hall}}

\subsection{A gentle history of the quantum Hall effect}
The quantum Hall effect is a phenomenon involving a collection of electrons restricted to move in a two-dimensional space and subject to a strong
magnetic field.\\
The classical Hall effect was discovered by Edwin Hall in 1879 \cite{Hall} and is a direct consequence of the motion of electrons in a magnetic field.
More precisely, it comes from the fact that the magnetic field causes electrons to move in circles. Let us recall quickly the calculation.
 This phenomena is known under the name cyclotron effect and is deduced from the equations
 of the motion for a particle
of mass $m$ and charge $-e$ in a $z$-directed magnetic field of intensity $B$:
\[{}
\left\{\begin{array}{l}\displaystyle
m{d^{2}\over dt^{2}}x=-eB{d\over dt}y \vspace{0.3cm}\\
\displaystyle m{d^{2}\over dt^{2}}y=eB{d\over dt}x,
 \end{array}\right.
\]
The general solution, $x(t)=x_{0}-r\sin(\omega_{B}t+\phi)$ and $y(t)=y_{0}+r\cos(\omega_{B}t+\phi)$, describes a circle.
 The parameters $x_{0}$, $y_{0}$, $r$ and $\phi${}
are chosen arbitrary, while  $\omega_{B}={eB\over m}$ is a linear function of $B$ and is called the cyclotron frequency. 
Taking into account an electric field $\vec E$ generating the current  together with a linear friction term modeled by the scattering time $\tau$, the motion equations become
\[{}
%m\left({d^{2}\over dt^{2}}x\atop {d^{2}\over dt^{2}}y\right)=eB\left({-d\over dt}y\atop {d^{2}\over dt^{2}}
m\left(d\over dt\right)\vec v=-e\vec v\wedge \vec B-{m\over\tau}\vec v-e\vec E.
\]
This model is known under the name of Drude model \cite{Drude,Drude2} and consists in considering the electrons as classical balls. 
Assuming that the velocity is constant, it can be written as:
\[{}
\left(\begin{array}{cc}m\over 2&-eB\\eB&m\over 2 \end{array}\right)\vec v=\vec E.
\]
 The current density $\vec J$ is related to the velocity by the equality
 $
 \vec J=-ne\vec v,
 $
 $n$ denoting the number of charged particles. Hence,
 $
 \vec E=\rho \vec J$
 where
 \[{}
 \rho={m\over e^{2}n\tau}\left(\begin{array}{cc}1&\omega_{B}\tau\\-\omega_{B}\tau&1\end{array}\right) \]
 denotes the conductivity.\\
 We see that there are two components to the resistivity: the off-diagonal component (Hall resistivity) $\rho_{xy}={m\omega_{B}\over e^{2}n}$, 
 which does not depend on $\tau$ but is linear in $B$,
 and the diagonal component (longitudinal resistivity) $\rho_{xx}={m\over e^{2}n\tau}$, which does not depend on $B$ and tends to $0$ when the scattering time $\tau${}
 tends to $\infty$.{}
   \begin{figure}[h]
	 \begin{center}
		 \includegraphics[width=10cm]{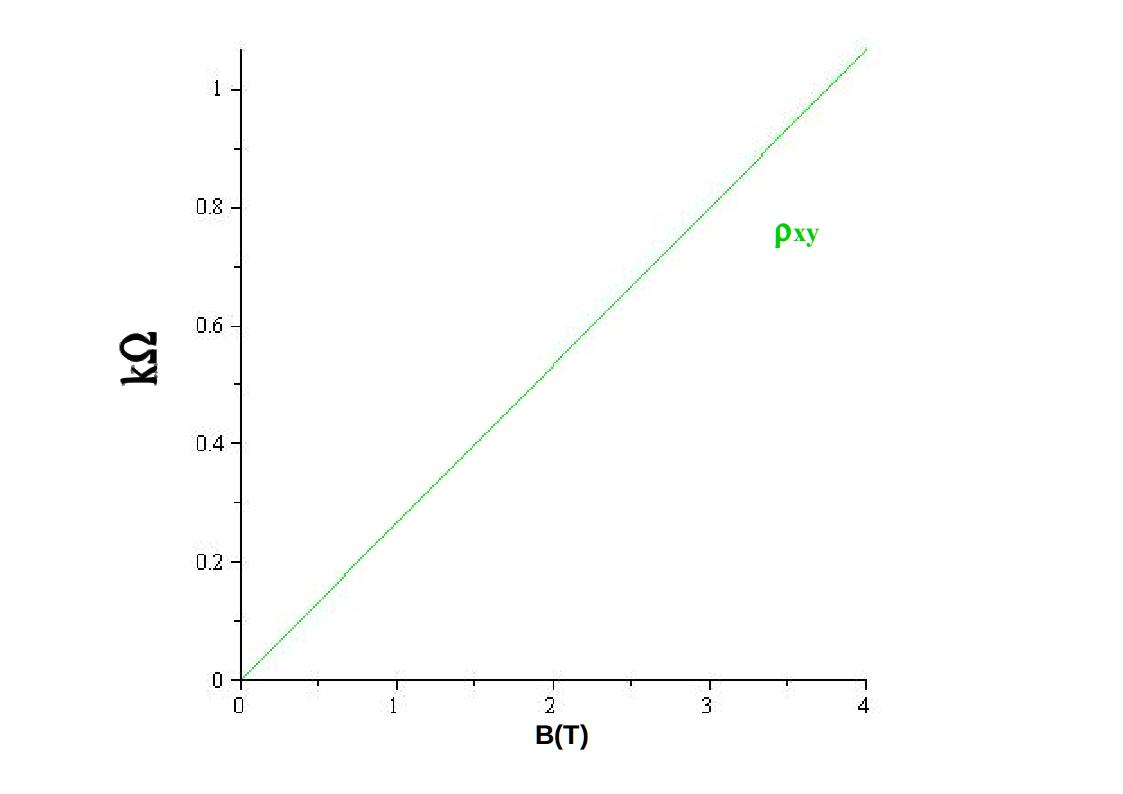}
	 \end{center}
	 \caption{Classical Hall effect}
 \end{figure} 
 In 1980, Von Klitzing et al. \cite{KDP} realized measurements of the Hall voltage of a two-dimensional electron gas with a 
 silicon metal-oxide-semiconductor field-effect transistor and showed the Hall resistivity has fixed values. 
 The exhibited phenomena is called the integer quantum
 Hall effect\footnote{Figure \ref{FigIQHE} was obtained by replacing the vertical scale $R_{H}$ by $\nu$ in \cite{Walet} (Figure 4.1, Section 4.4).
 This figure is under  
 \href{	https://creativecommons.org/licenses/by-nc-sa/3.0/}{Creative Commons Attribution-Noncommercial-Share Alike 3.0 Generic License} }.
  \begin{figure}[h]
	 \begin{center}
		 \includegraphics[width=10cm]{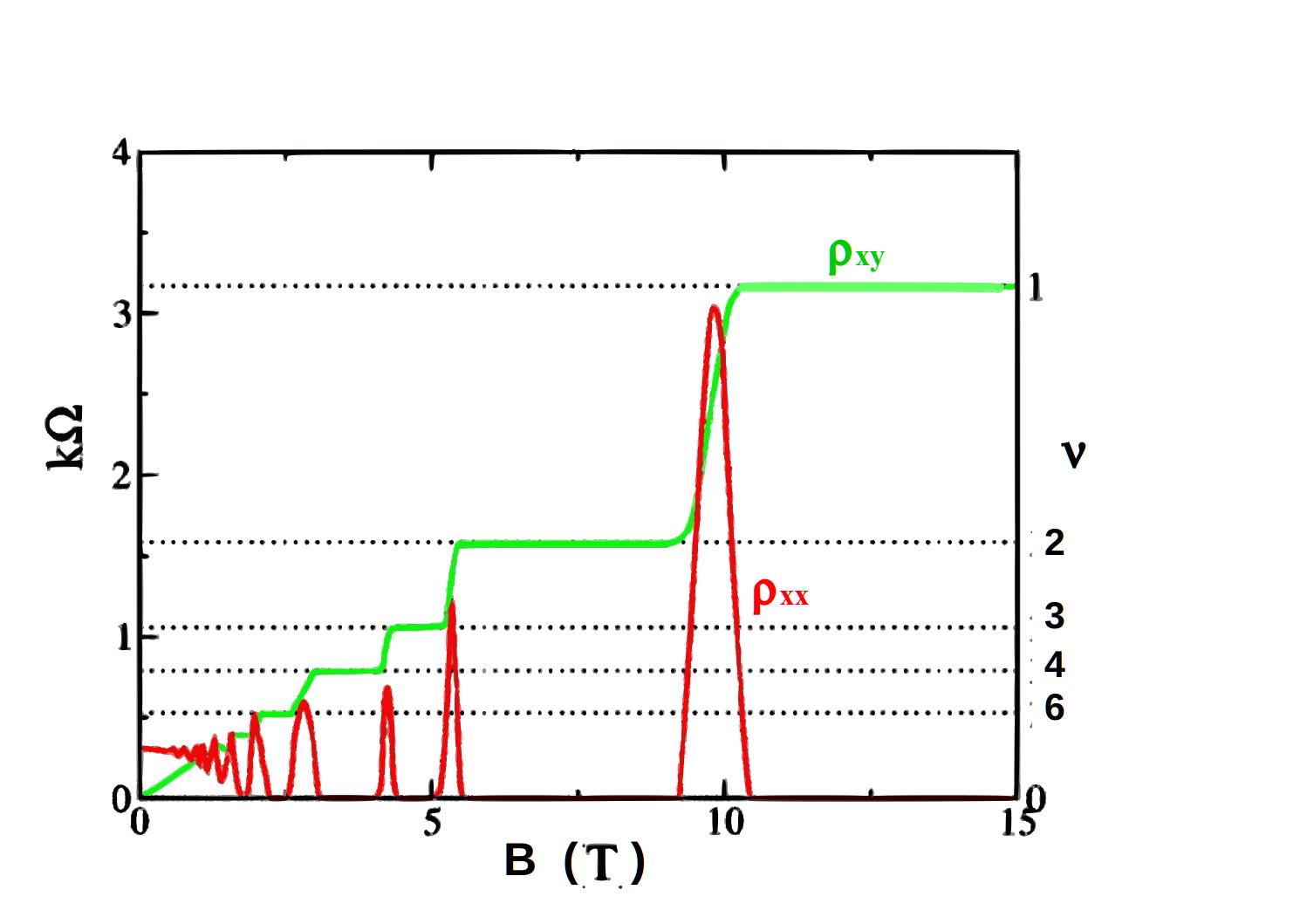}
	 \end{center}
	 \caption{Integral quantum Hall effect\label{FigIQHE}}
 \end{figure} 
Both the Hall resistivity and longitudinal resistivity have a behavior which highlights a quantum phenomena at the mesoscopic scale.
 The Hall resistivity is no longer a linear function of $B$ but sits on a plateau for a range of magnetic field before jumping 
 to the next one. These plateaus are centered on a values $B_{\nu}=r_{q}{n\over \nu}$, where $r_{q}={2\pi\hbar\over e}$ is the quantum resistivity,
  depending on a parameter $\nu\in\mathbb Z$ and
 the Hall resistivity takes the values
 $
 \rho_{xy}=\frac{r_{q}}{\nu}.
 $
 The longitudinal resistivity vanishes when $\rho_{xy}$ sits on a plateau and spikes when $\rho_{xy}$ jumps to the next one.
 
\begin{figure}[h]
	 \begin{center}
		 \includegraphics[width=10cm]{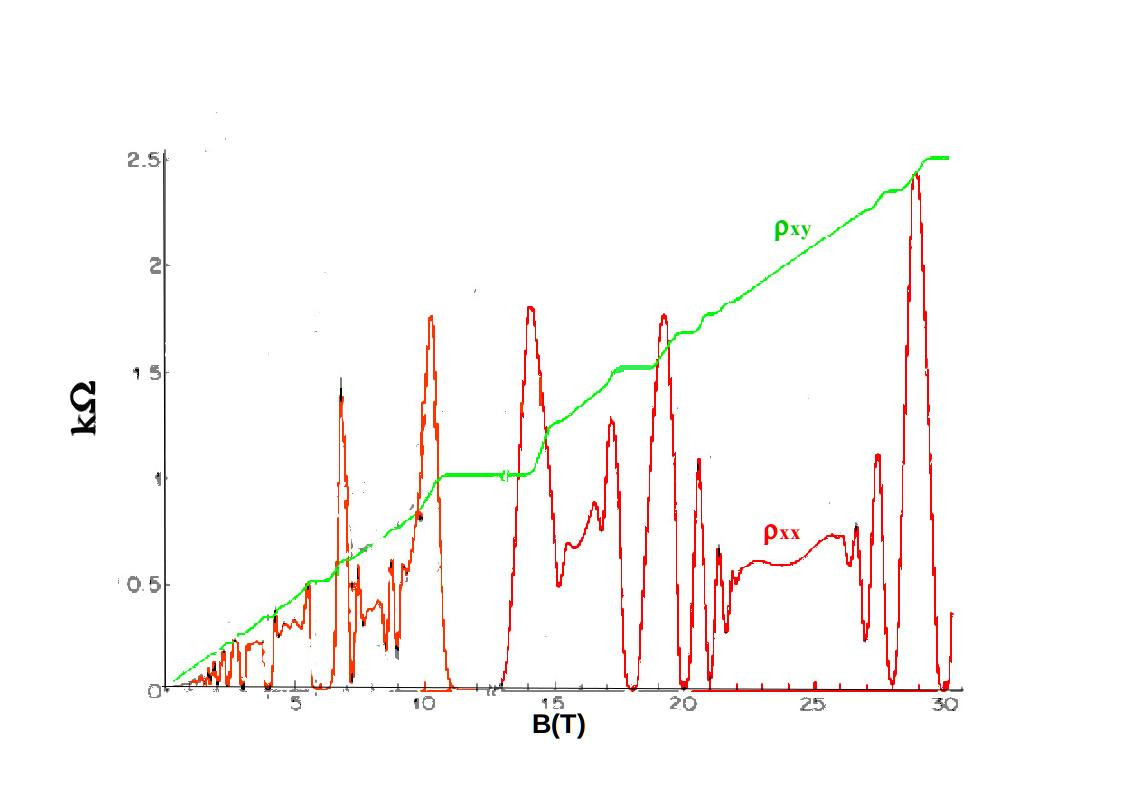}
	 \end{center}
	 \caption{Fractional quantum Hall effect \label{FigFQHE}}
 \end{figure}

 The fractional quantum Hall effect was discovered by Tsui et al \cite{TSG}. 
 They observed that, as the disorder is decreased, the integer Hall
 plateaus become less prominent and other plateaus emerge for fractional values of $\nu${}
 \footnote{
	 Figure \ref{FigFQHE} have been drawn by modifying a picture from \cite{WESTGE}.
	 }.{}
The difference between the integer quantum Hall effect 
 and the fractional quantum Hall effect is that, to explain the second, physicists need to take into account the interactions between the particles.
 The interaction between the electrons make the problem interesting from a mathematical point of view. 
 The theoretical approach was pioneered by Laughlin \cite{Laughlin}  for
 $\nu=\frac1{2m+1}$. Since the Hamiltonian is very difficult to diagonalize, 
 he proposed directly
  a wave function fulfilling several properties motivated by physical insight. 
  The Laughlin wave function overlaps more than $99\%$ with the true ground state.
   From the observations of Tsui et al and the work of Laughlin, more than 80 plateaus have been observed for various filling fractions. 
   The description of the wavefunctions is one of the challenges of the study. It is in this context that Jack polynomials appear.

 \subsection{Quantum Hall wavefunctions}
The fractional quantum Hall effect  appears in many configurations of the gas. 
Indeed, the Hall voltage can be generated by the motion of the 
particles but also by quasiparticles or quasiholes. Quasiparticles and quasiholes are virtual particles that occur when the matter behaves as 
if contained different weakly interacting particles. The charges of these virtual particles are  fractions of the electron charge and their masses are 
also different. But, in all the cases, for  fermion gases, the wave function must take the form
\[{}
\Phi(z_{1},\dots,z_{N})=\phi(z_{1},\cdots, z_{N})\prod_{i<j}(z_{i}-z_{j})\exp\left\{-\sum_{i=1}^{N}{-|z_{i}|^{2}\over 4\ell_{B}^{2}}\right\},
\]
where $\phi$ is a (polynomial) symmetric function, and $\ell_{B}=\sqrt{\hbar\over e B}$ is the magnetic length, which is a characteristic length scale 
governing quantum phenomena in a magnetic field. This expression is obtained assuming that the system is in the lowest Landau level, the single
particle wave functions take the form $\phi(z)=z^{m}e^{-|z|^{2}\over 4\ell^{2}_{B}}$, and that all the particles play the same role. This last condition
is a quite puzzling point. Indeed, since the particles are placed in a finite portion of the plane, they can not play the same role because the interactions must 
take into account the distance between the particles and the sides of the sample. Hence, the symmetry comes from an approximation
 when assuming
that the number of particles tends to infinity. The Haldane approach \cite{Haldane} for this theory consists in placing the 
particles on a sphere.
 The position of a particle is specified by spinor coordinates $u=\cos(\frac12\theta)e^{i\frac12\psi}$ and $v=\sin(\frac12\theta)e^{i\frac12\psi}$. When the radius tends to infinity the two approaches coincide and the wavefunctions in the spherical geometry are used to compute approximation for the plane geometry via stereographic projection.
 All the operators and wave function introduced by Haldane have been translated by this way in the plane geometry. 
 In what follow, we consider the Haldane point of view after stereographic projection.

The Laughlin wave function \cite{Laughlin} models the simplest FQH states which occurs for $\nu=\frac1m$. This wave function is given by
\begin{equation}\label{LaughlinState}
\phi_{Laughlin}^{m}(z_{1},\dots,z_{N}):=\prod_{i<j}(z_{i}-z_{j})^{2m}.
\end{equation}
Notice that $\phi_{Laughlin}$ can be seen as the stereographic projection of Haldane wavefunction 
stated in terms of spinor coordinates by
$$\phi_{Haldane}^{m}=\prod_{i<j}(u_{i}v_{j}-u_{j}v_{i})^{2m}.$$
In the Laughlin states no quasi-particle or quasihole  are involved. 
From a mathematical point of view the absence of quasi-particle and quasi-hole is
interpreted in terms of  differential operators as follows:
We consider the  operators $E_{n}:=\sum_{i}z_{i}^{n}{\partial\over\partial z_{i}}$ and 
 we set $L^{+}=E_{0}$ and $L^{-}=N_{\phi}\sum_{i}z_{i}-E_{2}$ where $N_{\phi}=2{\deg\phi\over N}$. The parameter $N_{\phi}$ is interpreted 
 in the spherical geometry by the fact that the sphere surrounds a monopole with charge $N_{\phi}$.
 The absence of quasiparticle is characterized by $L^{+}\phi=0$ (HW: \emph{highest weight} condition) 
 while the absence of quasihole is characterized by $L^{-}\phi=0$ (LW: \emph{lowest weight} condition). Noticing that $[E_{m},E_{n}]=(m-n)E_{n+m-1}$, we find
 that if $P$ is a polynomial satisfying the HW and LW conditions 
 then $E_{1}P=\frac12[E_{2},E_{0}]P=-\frac12E_{0}E_{2}P=\frac12N_{\phi}NP=\deg(P)P$, that is
 $P$ is an homogeneous polynomial. The HW condition means that the polynomial is invariant by translations. A fast computation shows that
 $\phi_{Laughlin}$ is both a HW and a LW state.\\
 Other interesting wavefunctions have been  exhibited. For instance, the Moore-Read (Pfaffian) state \cite{MR} is
 \begin{equation}\label{MRStates}
	 \phi^{0}_{MR}=\prod_{i<j}(z_{i}-z_{j})\mathrm{Pf}\left(1\over z_{i}-z_{j}\right),
 \end{equation}
 where  $\mathrm{Pf}$ denotes the Pfaffian.
 Surprisingly it describes the FQH for $\nu=1$. To understand the difference with the integer quantum Hall effect, physicists introduced two
 values $k$ and $r$ such that $\nu={k\over r}$. The parameter $k$ means that the function vanishes for $k+1$ particles together but not for $k$ 
 and the parameter $r$ is the order of the zeros. For instance, in the Laughlin state we have $r=2m$ and $k=1$, while for the Moore-Read state we have $r=2$ and $k=2$.
 In \cite{BH1,BH2}, Bernevig and Haldane showed how to associate to each Hall state an occupation number configuration. The occupation number configuration 
 is a vector $\mathbf n_{\phi}$ such that $\mathbf n_{\phi}[i]$ is the number of particles in the $i$th 
 lowest Landau level orbital ($i\geq 0$). 
 For a Laughlin state $\phi_{Laughlin}^{m}$ we have $\mathbf n_{\phi^{m}_{Laughlin}}=[1,0^{2m-1},1,0^{2m-1},\dots]$. For the Moore-Read state we have
 $\mathbf n_{\phi^{0}_{MR}}=[2,0,2,0,2,\dots]$.  The number $N_{\phi}$ is the greatest integer $i$ such that $n_{\phi}[i]\neq 0$. 
 Instead to use the occupation number configuration, we will use a decreasing partition $\lambda_{\phi}$ such that the multiplicity of the part $i$ in $\lambda_{\phi}$ equals $\mathbf n_{\phi}[i]$.
 For instance $\lambda_{\phi^{m}_{Laughlin}}=[(N-1)m,\dots,m,0]$ and $\lambda_{\phi^{0}_{MR}}=\left[2({N\over 2}-1),2({N\over 2}-1),\dots,4,4,2,2,0,0\right]$ ($N$ needs to be even for Moore-Read states).
 We see that $N_{\phi}$ is the biggest part in $\lambda_{\phi}$.\\
 Reader interested by fractional quantum Hall theory can refer to \cite{Tong} for a complete picture on the topic.

\subsection{FQHT and Jack polynomials}
 
 Some of the trial wave functions proposed to describe the FQHE are Jack functions. This is the case of the simplest one,
 \begin{equation}\label{LaughlinisJack}
 \phi^{m}_{Laughlin}(z_{1},\dots,z_{N})\displaystyle{\mathop=^{(*)}}
J^{\left(-2\over 2m-1\right)}_{[2(N-1)m,2(N-2)m,\dots,0]}(z_{1},\dots,z_{N}).
 \end{equation}
 This was first noticed by Bernevig and Haldane \cite{BH1}. They obtained this equality by proving that $\phi^{m}_{Laughlin}$ is an eigenstate
 of the Laplace-Beltrami operator
 \begin{equation}\label{LBO}
 \mathcal H^{(\alpha)}=\sum_{i}\left(z_{i}{\partial\over\partial z_{i}}\right)^{2}+
 \frac1\alpha\sum_{i<j}{z_{i}+z_{j}\over z_{i}-z_{j}}\left(z_{i}{\partial\over\partial z_{j}}-z_{j}{\partial\over\partial z_{i}}\right).
 \end{equation}
 This is particularly interesting to remark that the main argument of the proof comes from clustering properties.
  Indeed, the Laughlin wavefunction, considered as a polynomial in $z_{i}$ (for some $i\in\{1,\dots,N\}$) has a multiplicity $2m$ root at $z_{i}=z_{j}$ 
  for any $j\neq i$. So it vanishes under the action of the operator $D_{i}^{L,2m}$ where $D_{i}^{L,r}={\partial\over\partial z_{i}}-r\sum_{j\neq i}\frac1{z_{i}-z_{j}}$.
  Since $\phi_{Lauglin}^{m}$ is in the kernel of 
  $\sum_{i}z_{i}D_{i}^{L,1}z_{i}D_{i}^{L,r}=\mathcal H^{\left(-\frac2{2m-1}\right)}-{m\over 6}N(N-1)(N+1+6m(N-1))$,
  it is an eigenfunction of $\mathcal H^{\left(-\frac2{2m-1}\right)}$. Hence, $\phi_{Laughlin}^{m}$ is identified with 
  $J^{\left(-2\over 2m-1\right)}_{[2(N-1)m,2(N-2)m,\dots,0]}(z_{1},\dots,z_{N})$ by considering its dominant monomial.

In the same paper \cite{BH1}, a similar (but a little more complicated) reasoning is used to study the Moore-Read state described in \cite{MR}: $\phi^{0}_{MR}$ (\ref{MRStates}).
They proved that
\begin{equation}\label{MRisJack}
	\phi^{0}_{MR}(z_{1},\dots,z_{N})\displaystyle{\mathop=^{(*)}}
J^{(-3)}_{\left[2({N\over 2}-1),2({N\over 2}-1),\dots,4,4,2,2,0,0\right]}(z_{1},\dots,z_{N}).
\end{equation}
Other examples are treated in \cite{BH1}. For instance, the $Z_{p}$ parafermionic states
\begin{equation}\label{RRState}
	\phi_{RR}^{0}=\mathcal S\left(\prod_{k=1}^{N}\prod_{(k-1)N'\leq i<j\leq kN'}(z_{i}-z_{j})^{2}\right){}
	\displaystyle{\mathop=^{(*)}}
J^{-(p+1)}_{[(2(N'-1))^{p},\dots,2^{p},0^{p}]}(z_{1},\dots,z_{N}),
\end{equation}
where $N=pN'$ and $\mathcal S$ denotes the symmetrizing operator. 
This example generalizes (\ref{MRisJack}) and is a special case of a Read-Rezayi state for $\nu={p\over 2}$ \cite{RR}.
A more complicated example is involved at $\nu={2\over 5}$ and refereed to as "Gaffnian" \cite{SRCB}. This wavefunction is also proved to be
a Jack polynomial \cite{BH1}
\begin{equation}\label{GState}
	\phi_{G}(z_{1},\dots,z_{N})\displaystyle{\mathop=^{(*)}}
J^{\left(-3\over 4\right)}_{[3(N'-1),3(N'-1),3(N'-2),3(N'-2),\cdots,3,3,0,0}(z_{1},\dots,z_{N}),
\end{equation}
with $N=2N'$.\\
In the aim to provide tools for the understanding of FQH states, Bernevig and Haldane investigated clustering properties of Jack polynomials, \cite{BH2}.
In particular, they exhibited a family of highest weight Jack polynomials in $N$ variables that vanish when $s$ distinct clusters of $k+1$ particles are formed.
The corresponding partitions depends on $4$ parameters (the parameter $\beta$ depends  on $N$ and on $3$ other parameters, and is implicit in \cite{BH2}) $\lambda_{k,r,s}^{\beta}=[(\beta r+s(r-1)+1)^{k},\dots,(s(r-1)+1)^{k},0^{n_{0}}]$ with $n_{0}=(k+1)s-1$ and $N=\beta k+n_{0}$.
Notice that, in this case, the flux (i.e. the maximal degree in each variable) equals
\begin{equation*}
	N_{\phi}=\beta r+s(r-1)+1={r\over k}(N-k-(k+1)(s-1))+(r-1)(s-1).
\end{equation*}
 Bernevig and Haldane investigated three kinds of clustering properties that occur when $k+1$ and $s-1$ are coprime:
 \begin{enumerate}
	 \item {\bf First clustering property} They considered $s-1$ clusters of $k+1$ particles 
	 $Z_{1}=z_{1}=\cdots=z_{k+1}$, $Z_{2}=z_{k+1}=\cdots=z_{2(k+1)}$,$\dots$,$Z_{s-1}=z_{(s-2)(k+1)+1}=\cdots=z_{(s-1)(k+1)}$, together with 
	 $k$ particle cluster $Z_{F}=z_{(s-1)(k+1)}=\cdots=z_{s(k+1)-1}$.
	  The other particles (variables) remain free. For such a specialization,
	 the Jack polynomial $J^{-{k+1\over r-1}}_{\lambda_{k,r,s}^{\beta}}((k+1)(Z_{1}+\cdots+Z_{s-1})+kZ_{F}+z_{s(k+1)}+\cdots+z_{N})$ behaves as $\displaystyle\prod_{i=s(k+1)^{N}}\left(Z_{F}-z_{i}\right)^{r}$ when 
	 each $z_{i}$ ($i=s(k+1),\dots,N$) tends to $Z_{F}$. For instance, we have
	 \[{}
	 J^{(-2)}_{53}(2Z_{1}+Z_{F}+z_{3}+z_{4})= \left( {\it Z_F}-{\it z_4}
 \right) ^{2} \left( {\it Z_F}-{\it z_3} \right) ^{2}P(Z_{1},Z_{F},z_{3},z_{4})\]
 with \begin{multline*}
 P(Z_{1},Z_{F},z_{3},z_{4})=
	 144\, \left( {\it z_3}-{\it z_4} \right) ^{2} \\
	 \times  \left( {\it z_3}\,{
\it z_4}+{\it Z_F}\,{\it z_4}+{\it Z_F}\,{\it z_3}-2\,{\it Z_1}\,{\it z_4}-2
\,{\it Z_1}\,{\it z_3}-2\,{\it Z_1}\,{\it Z_F}+3\,{{\it Z_1}}^{2} \right).
	 \end{multline*}
	 \item {\bf Second clustering property} They considered a  cluster of $n_{0}=(k+1)s-1$ particles $z_{1}=\cdots=z_{(k+1)s-1}=Z$.
	 The  Jack polynomial $J_{\lambda_{k,r,s}}^{\left(-{k+1\over r-1}\right)}(n_{0}Z+z_{n_{0}+1}+\cdots+z_{N})$ behaves as
	 $\displaystyle\prod_{i=s(k+1)}^{N}(Z-z_{i})^{(r-1)s+1}$ when each $z_{i}$ tends to $Z$. More specifically, for highest weight Jack polynomials, one has the 
	 following explicit formula:
\begin{equation}\label{SCC}
	J_{\lambda_{k,r,s}^{\beta}}^{\left(-{k+1\over r-1}\right)}(n_{0}Z+z_{n_{0}+1}+\cdots+z_{N})\displaystyle{\mathop=^{(*)}}
\prod_{i=s(k+1)}^{N}(Z-z_{i})^{(r-1)s+1}
	J_{\lambda_{k,r,1}^{\beta-1}}^{\left(-{k+1\over r-1}\right)}(z_{n_{0}+1}+\cdots+z_{N}).
\end{equation}
For instance,
\[{}
J_{53}^{(-2)}(3Z+z_{4}+z_{5})=-144(Z-z_{4})^{3}(Z-z_{5})^{3}J_{2}^{(-2)}(z_{4}+z_{5}).
\]
	 \item {\bf Third clustering property} It is obtained by forming  $s-1$ clusters of $2k+1$ particles $Z_{1}=z_{1}=\cdots=z_{2k+1}$,$\dots$,{}
	 $Z_{s-1}=z_{(s-2)(2k+1)+1}=\cdots=z_{(s-1)(2k+1)}$. A highest weight Jack $J_{\lambda^{\beta}_{k,r,s}}$ satisfies
	 \begin{multline}
\displaystyle J_{\lambda^{\beta}_{k,r,s}}^{\left(-{k+1\over r-1}\right)}((2k+1)(Z_{1}+\cdots+Z_{s-1})+z_{(s-1)(2k+1)+1}+\cdots+z_{N}) \\
\displaystyle{\mathop=^{(*)}}
\prod_{1\leq i<j\leq s-1}(Z_{i}-Z_{j})^{k(3r-2)}\displaystyle\prod_{i=1}^{s-1}\prod_{\ell=(s-1)(2k+1)+1}^{N}(Z_{i}-z_{\ell})^{2r-1} \\
\times J_{\lambda^{\beta-s+1}_{k,r}}(z_{(s-1)(2k+1)+1}+\cdots+z_{N}).
	 \end{multline}
	 For instance,
	 \[{}
	 J_{64}^{\left(-2\right)}(3(Z_{1}+Z_{2})+z_{7})=-3456(Z_{1}-Z_{2})^{4}(Z_{1}-z_{7})^{3}(Z_{2}-z_{7})^{3}.
	 \]
 \end{enumerate}
 The aim of our paper is to show how the material described in \cite{DL3} can help in this context. In particular, we focus on the
 second clustering property for HW polynomials. \\
 To be more complete, the wavefunctions are not all Jack polynomials but many of them can be obtained from Jack polynomials by acting by an operator
 modeling the adding of a quasiparticle or a quasihole (see e.g. \cite{BH3}).
 \subsection{The interest of shifted Macdonald polynomials}
 In \cite{BH1} (i.e. for $s=1$), Bernevig and Haldane proved the clustering properties on HW Jack polynomials using a result of 
 B. Feigin et al \cite{FJMM} together with 
 Lassalle binomial formulas for Jack polynomials \cite{Lassalle}. Lassalle binomial formula are used to describe the action
 of the operator $L^{+}$ on a Jack polynomial. When $s>1$, the partitions do not fulfill some admissibility conditions of B. Feigin et al. \cite{FJMM},
 and so, the equations are just conjectured  from extensive numerical computations. For the purpose of manipulating these identities properly, we must leave the framework of homogeneous Jack polynomials.
 First, clustering properties deal with vanishing properties. So shifted Jack polynomials should be more appropriate for these problems.
 Nevertheless, the multiplicities of the roots of the polynomials are difficult to manage. The idea consists in $(q,t)$-deforming these identities
 in such a way that they involve products of distinct factors. For instance, a factor $(z_{i}-z_{j})^{n}$ should become $(z_{i}-z_{j})(z_{i}-qz_{j})\cdots (z_{i}-q^{n-1}z_{j})$.
 With such a deformation, it is also easier to manipulate the eigenspaces which are smaller (see \cite{BL} for the example related 
 to $\phi_{Laughlin}$).  In consequence, we follow the strategy initiated in \cite{DL3}, which consists to manipulate shifted Macdonald polynomials in the aim to prove the identities.
 The recipe is as follows:
 \begin{itemize}
	 \item We find a Macdonald version of the conjecture and we state it in terms of vanishing properties.
	 \item We prove that the Macdonald polynomial involved is a highest weight polynomial (i.e. in the kernel of a $q$-deformation of $L^{+}$). 
	 When it is possible, this property comes from \cite{FM} (Macdonald version of \cite{FJMM}), while for the other cases we apply the results of \cite{JL}, which are based on the Lassalle binomial formula for Macdonald polynomials \cite{Lassalle}.
	 \item In this case, the shifted Macdonald polynomial equals the homogeneous Macdonald polynomials.
	 \item We deduce the equality from vanishing properties of the shifted Macdonald polynomial.
	  \item We recover the identity on Jack by sending $q$ to $1$.
 \end{itemize}
 Notice that in \cite{JL}, one of the authors with Thierry Jolicoeur found some families of polynomials which 
 have not been considered in \cite{BH2}.
 Indeed, Bernevig and Haldane missed that the family $\lambda_{k,r,s}^{\beta}$ can be extended by adding a parameter corresponding to the multiplicity of the largest part which can be smaller than $k$.
 Also, some other Macdonald polynomials do not specialize to a Jack for the considered specialization of $(q,t)$.\\
  We detail all of that in the next sections.
\section{Factorizations for generic $(q,t)$\label{FactGen}}
\subsection{Saturated partitions}
We set $N>0$. We say that a  partition $\lambda=[\lambda_{1},\dots,\lambda_{N}]$ is {\it saturated} if $\lambda_{N}>0$.
\begin{proposition}\label{saturate}
If $\lambda=[\lambda_{1},\dots,\lambda_{N}]$ is saturated, then:
	\begin{enumerate}
		\item $P_{\lambda}(x_{1},\dots,x_{N};q,t)=(x_{1}\cdots x_{N})^{\lambda_{N}}
		P_{[\lambda_{1}-\lambda_{N},\dots,\lambda_{N-1}-\lambda_{N},0]}(x_{1},\dots,x_{N};q,t).$
		\item $MS_{\lambda}(x_{1},\dots,x_{N};q,t)\displaystyle\mathop=^{(*)}\prod_{k=0}^{\lambda_{N}-1}\prod_{i=1}^{N}(x_{i}-q^{k})
		MS_{[\lambda_{1}-\lambda_{N},\dots,\lambda_{N-1}-\lambda_{N},0]}(q^{-\lambda_{N}}x_{1},\dots,q^{-\lambda_{N}}x_{N};q,t).$
	\end{enumerate}
\end{proposition}
Recall that $\displaystyle\mathop=^{(*)}$ means that the equality holds up to a scalar factor.

{\it Proof.}
Recall the affine step for non symmetric Macdonald polynomials (resp. Shifted Macdonald polynomials)%
\[
E_{v\Phi}=E_{v}\tau x_{N}\ \ \ \  \left(\mbox{resp.}\ M_{v\Phi}=M_{v}\tau\left(  x_{N}-1\right)\right)
\]
and $v\Phi=\left[  v\left[  2\right]  ,v\left[  3\right]  ,\ldots,v\left[
N\right]  ,v\left[  1\right]  +1\right]  $. If we apply this step $N$ times, we obtain%
\[
E_{\left[  v\left[  1\right]  +1,v\left[  2\right]  +1,\ldots,v\left[
N\right]  +1\right]  }=\prod\limits_{i=1}^{N}x_{i}
E_{v}\left(  \frac{x_{1}}{q},\frac{x_{2}}{q},\ldots\frac{x_{N}}{q}\right)  ;
\]
respectively,
\[
M_{\left[  v\left[  1\right]  +1,v\left[  2\right]  +1,\ldots,v\left[
N\right]  +1\right]  }=\prod\limits_{i=1}^{N}\left(  x_{i}-1\right)
M_{v}\left(  \frac{x_{1}}{q},\frac{x_{2}}{q},\ldots\frac{x_{N}}{q}\right) .
\]
Hence, if we apply the affine step $N$ times again, we obtain
\[
E_{\left[  v\left[  1\right]  +2,v\left[  2\right]  +2,\ldots,v\left[
N\right]  +2\right]  }=\prod\limits_{i=1}^{N}{x_{i}^{2}\over q}  
  E_{v}\left(  \frac{x_{1}}{q^{2}},\frac{x_{2}}{q^{2}%
},\ldots\frac{x_{N}}{q^{2}}\right);
\]
respectively,
\[
M_{\left[  v\left[  1\right]  +2,v\left[  2\right]  +2,\ldots,v\left[
N\right]  +2\right]  }=\prod\limits_{i=1}^{N}\left(  x_{i}-1\right)  \left(
\frac{x_{i}}{q}-1\right)  M_{v}\left(  \frac{x_{1}}{q^{2}},\frac{x_{2}}{q^{2}%
},\ldots\frac{x_{N}}{q^{2}}\right)  .
\]
By induction, starting with $v=\left[  \lambda_{1}-\lambda_{N},\lambda
_{2}-\lambda_{N},\ldots,\lambda_{N-1}-\lambda_{N},0\right]  $ and applying the
affine step $N\lambda_{N}$ times, one finds
\[
E_{\lambda}\mathop=^{(*)}\prod_{i=1}^{N}{x_{i}^{\lambda_N}}  
  E_{v}\left(  \frac{x_{1}}{q^{\lambda_{N}}},\frac{x_{2}}{q^{\lambda_{N}}%
},\ldots\frac{x_{N}}{q^{\lambda_{N}}}\right),\] and \[
M_{\lambda}\mathop=\prod\limits_{i=1}^{N}\prod_{k=0}^{\lambda_{N}-1}\left({x_{i}^{\lambda_N}\over q^{k}}-1\right)  
  M_{v}\left(  \frac{x_{1}}{q^{\lambda_{N}}},\frac{x_{2}}{q^{\lambda_{N}}%
},\ldots\frac{x_{N}}{q^{\lambda_{N}}}\right).
\]
Since the polynomials $\prod\limits_{i=1}^{N}{x_{i}^{\lambda_N}}$ and $\prod\limits_{i=1}^{N}\prod\limits_{k=0}^{\lambda_{N}-1}
\left({x_{i}^{\lambda_N}\over q^{k}}-1\right) $ are symmetric, they commute
with the action of the symmetrizing operator $\mathcal{S}$ and the result is obtained by applying the symmetrizing operator to $E_{\lambda}$ and $M_{\lambda}$. \qed %$\Box$

	 \begin{remark}\rm
		  Notice that one has an alternative proof for the second result.
		   One has to examine the vanishing properties of \begin{equation}\label{prod2}\prod_{k=0}^{\lambda_{N}-1}
	 \prod_{i=1}^{N}(x_{i}-q^{k})MS_{[\lambda_{1}-\lambda_{N},\dots,\lambda_{1}-\lambda_{N-1},0]}
	 (q^{-\lambda_{N}}\mathbb X_{N};q,t).\end{equation} 
	 
	 Let $\mu\neq\lambda$, with $|\mu|\leq|\lambda|$.
	 If $[\lambda_{N},\dots,\lambda_{N}]\subset\mu$ then  the vanishing properties of $ MS_{[\lambda_{1}-\lambda_{N},\dots,\lambda_{1}-\lambda_{N-1},0]}$ gives
\begin{multline*}
	 MS_{[\lambda_{1}-\lambda_{N},\dots,\lambda_{1}-\lambda_{N-1},0]}(q^{-\lambda_{N}}q^{\mu_{1}}t^{N-1}+\cdots+q^{-\lambda_{N}}q^{\mu_{N-1}};q,t)=\\
	 =MS_{[\lambda_{1}-\lambda_{N},\dots,\lambda_{1}-\lambda_{N-1},0]}(\langle \mu_{1}-\lambda_{N},\dots,\mu_{N}-\lambda_{N}\rangle)=0,
\end{multline*}
	 because $[ \mu_{1}-\lambda_{N},\dots,\mu_{N}-\lambda_{N}]\neq [\lambda_{1}-\lambda_{N},\dots,\lambda_{1}-\lambda_{N-1},0]$ and 
	 $|[ \mu_{1}-\lambda_{N},\dots,\mu_{N}-\lambda_{N}]|\leq |[\lambda_{1}-\lambda_{N},\dots,\lambda_{1}-\lambda_{N-1},0]|$.\\
	 If $[\lambda_{N},\dots,\lambda_{N}]\not\subset\mu$ then this means that $\mu_{N}<\lambda_{N}$. Then, the factor $(x_{N}-q^{\mu_{N}})$ in (\ref{prod2})
	 vanishes for $x_{N}=q^{\mu_{N}}$. This proves that the two polynomials have the same vanishing properties and so, that they are equal.
	 \end{remark}
\subsection{Standard specializations}
If $\lambda=[\lambda_{1},\dots,\lambda_{N-k},0^{k}]$, the \emph{$\lambda$-standard specialization} consists in setting $x_{N-k+1}=t^{k-1}, 
x_{N-k+2}=t^{k-2}, \dots, x_{N}=1$.
\begin{proposition}\label{standard}
\begin{multline*}
	MS_{[\lambda_{1},\dots,\lambda_{N-k},0^{k}]}\left(\mathbb X_{N-k}+{1-t^{k}\over 1-t};q,t\right)\displaystyle\mathop=^{(*)}{}
	MS_{[\lambda_{1},\dots,\lambda_{N-k}]}(t^{-k}\mathbb X_{N-k};q,t)\mathop=^{(*)} \\\displaystyle\mathop=^{(*)}{}\
	\prod_{j=0}^{\lambda_{N-k}-1}\prod_{i=1}^{N-k}(x_{i}-t^{k}q^{j})MS_{[\lambda_{1}-\lambda_{N-k},\dots,\lambda_{N-k-1}-\lambda_{N-k},0]}
	\left(q^{-\lambda_{N-k}}t^{-k}\mathbb X_{N-k};q,t\right).
\end{multline*}
\end{proposition}
{\it Proof.} 
Consider the polynomial $$\mathcal P(x_{1},\dots,x_{N-k})=MS_{[\lambda_{1},\dots,\lambda_{N-k},0^{k}]}\left(t^{k}\mathbb X_{N-k}t^{k}+{1-t^{k}\over 1-t};q,t\right).$${}
This polynomial vanishes for $[x_{1},\cdots,x_{N-k}]=[q^{\mu_{1}}t^{N-k-1},\dots,q^{\mu_{N-k}}t^{0}]$, 
for any $\mu\neq\lambda$. Moreover, $|\mu|\geq |\lambda|$,{}
 since
$$\langle \lambda_{1},\dots,\lambda_{N-k},0^{k}\rangle=[q^{\lambda_{1}}t^{N-1},\dots,q^{\lambda_{N-k}}t^{N-k},t^{k-1},\dots,1].$$ So 
$\mathcal P(x_{1},\dots,x_{N-k})$ has the same vanishing properties as $MS_{[\lambda_{1},\dots,\lambda_{N-k}]}$. This proves the first equality.
 The second equality is a  direct consequence of Proposition \ref{saturate}.\qed
 \begin{example}\rm 
 We illustrate the principle of the proof with $\lambda=[3,2,0,0,0]$. The vanishing properties of $MS_{3200}$ implies that  $MS_{3200}(x_{1},x_{2},t^{2},t,1)$ vanishes for the following values of
 $(x_{1},x_{2})$: $(q^{5}t^{4},t^{3})$,  $(q^{4}t^{4},qt^{3})$, $(q^{4}t^{4},t^{3})$, $(q^{3}t^{4},qt^{3})$, $(q^{2}t^{4},q^{2}t^{3})$, $(q^{3}t^{4},t^{3})$, $(q^{2}t^{4},qt^{3})$,
 $(q^{2}t^{4},t^{3})$, $(qt^{4},qt^{3})$, $(qt^{4},t^{3})$ and $(t^{4},t^{3})$. Since, $MS_{3200}(x_{1},x_{2},t^{2},t,1)$ is a degree $5$ symmetric polynomial in two variables, these 
 vanishing properties completely characterize it up to a global factor. 
 Indeed, there are exactly $12$ independent symmetric functions of degree at most $5$ in $2$ variables. The basis of the space spanned by these functions
 is generated by the polynomials $MS_{50}(x_{1},x_{2};q,t)$, $MS_{41}(x_{1},x_{2};q,t)$, $MS_{32}(x_{1},x_{2};q,t),\  \dots,$ 
 $MS_{00}(x_{1},x_{2})$. The polynomial $MS_{3200}(x_{1},x_{2},t^{2},t,1)$ is symmetric and so, is a linear combination of the $12$ polynomials above.
 It follows that a series of $11$ vanishing properties is sufficient to produce a system of linear equations characterizing the coefficients of
 this combination.
\\
 Comparing to the vanishing properties characterizing $MS_{32}$:
 \[{}
 \begin{array}{|c|c|c|}
	 \hline \langle50\rangle&\langle41\rangle&\langle 32\rangle\\\ 
	 [q^{5}t,1]&[q^{4}t,q]&\times \\\hline
	 \langle40\rangle&\langle31\rangle&\langle22\rangle\\\ 
	 [q^{4}t,1]&[q^{3}t,q]&[q^{2}t,q^{2}]\\\hline\
	 \langle30\rangle&\langle21\rangle&\\\ 
	 [q^{3}t,1]&[q^{2}t,q]&\\\hline\
	 \langle20\rangle&\langle11\rangle&\\\ 
	 [q^{2}t,1]&[qt,q]&\\\hline\ 
	 \langle10\rangle&&\\\ 
	 [qt,1]&&\\\hline 
	 \langle00\rangle&&\\\
	 [t,1]&&\\\hline
 \end{array}
 \]
 we deduce that the polynomials $MS_{32000}(\mathbb X_{2}+t^{2}+t+1)$ and $MS_{32}(t^{-3}\mathbb X_{2})$ are proportional.\\ 
 These two polynomials have low degree and so, are easy to compute by the help of the Yang-Baxter graph. One finds 
 \[{}
 MS_{32}(\mathbb X_{2};q,t)\mathop=^{(*)} \left( {q}^{2}t+{q}^{2}-{\it x_1}-{\it x_2} \right)  \left( {\it x_2}-1
 \right)  \left( {\it x_1}-1 \right)  \left( -{\it x_2}+q \right) 
 \left( -{\it x_1}+q \right),{}
 \]
 and
 \[{}\begin{array}{l}
 MS_{3200}(\mathbb X_{2}+t^{2}+t+1;q,t)\displaystyle\mathop=^{(*)} \left( {t}^{3}-{\it x_2} \right)  \left( {t}^{3}-{\it x_1} \right) 
 \left( q{t}^{3}-{\it x_2} \right)  \left( q{t}^{3}-{\it x_1} \right) 
 \left( {q}^{2}{t}^{3}+{q}^{2}{t}^{4}-{\it x_1}-{\it x_2} \right). \end{array}
 \]
 \end{example}
 \begin{corollary}
	 Denoting by $\lfloor \lambda\rfloor_{i}$ the number of parts of $\lambda$ lower or equal to $i$, and by $m_{\lambda}$ the multiplicity of
	 the maximal part in $\lambda$, we have
	 \begin{equation*}
	 MS_{\lambda}\left(\mathbb X_{m_{\lambda}}+\lbag\lambda_{m_{\lambda}+1},\dots,\lambda_{N-1},\lambda_{N}\rbag\right)\mathop=^{(*)}
	 \prod_{j=0}^{\max \lambda-1}\prod_{i=1}^{m_{\lambda}}(x_{i}-q^{j}t^{\lfloor \lambda\rfloor_{j}}).
	 \end{equation*}
 \end{corollary}
 {\it Proof.} We prove the property by induction on $N+|\lambda|$. We have to consider two cases:
 \begin{enumerate}
	 \item If $\lambda_{N}>0$, then by Proposition \ref{saturate}, we have
	 \begin{multline}
	 MS_{\lambda}\left(\mathbb X_{m_{\lambda}}+\lbag\lambda_{m_{\lambda}+1},\dots,\lambda_{N-1},\lambda_{N}\rbag\right)\displaystyle\mathop=^{(*)}\displaystyle\prod_{j=0}^{\lambda_{N}-1}\prod_{i=1}^{m_{\lambda}}(x_{i}-q^{j})\times\\\times
	 MS_{[\lambda_{1}-\lambda_{N},\dots,\lambda_{N-1}-\lambda_{N},0]}\left(q^{-\lambda_{N}}\left(\mathbb X_{m_{\lambda}}+
	 \lbag\left[\lambda_{m_{\lambda}+1},\dots,\lambda_{N-1},\lambda_{N}\right]\rbag\right)\right).
	 \end{multline}
	 Setting $\tilde\lambda=[\lambda_{1}-\lambda_{N},\dots,\lambda_{N-1}-\lambda_{N},0]$, we have $\lbag\tilde\lambda\rbag={}
	 q^{-\lambda_{N}}\lbag\lambda_{m_{\lambda}+1},\dots,\lambda_{N-1},\lambda_{N}\rbag$ and by induction:
	 \begin{equation*}
	  MS_{\tilde\lambda}(q^{-\lambda_{N}}\mathbb X_{m_{\lambda}}+\lbag\tilde\lambda\rbag)\mathop=^{(*)}\prod_{j=0}^{\max \tilde\lambda-1}\prod_{i=1}^{m_{\tilde\lambda}}
	 (x_{i}-q^{j+\lambda_{N}}t^{\lfloor\tilde \lambda\rfloor_{j}}).
	 \end{equation*}
	 But $\lfloor\tilde \lambda\rfloor_{j}=\lfloor\lambda\rfloor_{j+\lambda_{N}}$. Hence,
	 \begin{multline}
	 \displaystyle	 \prod_{j=0}^{\lambda_{N}-1}\prod_{i=1}^{m_{\lambda}}(x_{i}-q^{j})\prod_{j=0}^{\max \tilde\lambda-1}\prod_{i=1}^{m_{\tilde\lambda}}
	 (x_{i}-q^{j+\lambda_{N}}t^{\lfloor\tilde \lambda\rfloor_{j}})=\\=\displaystyle
	 \prod_{j=0}^{\lambda_{N}-1}\prod_{i=1}^{m_{\lambda}}(x_{i}-q^{j})\prod_{j=\lambda_{N}}^{\max \lambda-1}\prod_{i=1}^{m_{\lambda}}
	 (x_{i}-q^{j}t^{\lfloor \lambda\rfloor_{j}}) =\displaystyle\prod_{j=0}^{\max \lambda-1}\prod_{i=1}^{m_{\lambda}}(x_{i}-q^{j}t^{\lfloor \lambda\rfloor_{j}}).
	 \end{multline}
	 as expected.
	 \item If $\lambda_{N}=0$, then we set $\lambda=[\lambda_{1},\dots,\lambda_{N-k},0^{k}]$ and, by Proposition \ref{standard}, we obtain
	\begin{equation*} MS_{[\lambda_{1},\dots,\lambda_{N-k},0^{k}]}\left(\mathbb X_{N-k}+{1-t^{k}\over 1-t};q,t\right)\displaystyle\mathop=^{(*)}{}
	MS_{[\lambda_{1},\dots,\lambda_{N-k}]}(t^{-k}\mathbb X_{N-k};q,t).\end{equation*}
	We conclude by applying the part 1 of the proof.
	%Setting $\tilde\lambda=[\lambda_{1},\dots,\lambda_{N-k}]$, one obtains by induction
	% \[{}\begin{array}{r}\displaystyle
	% MS_{\tilde\lambda}(x_{1}t^{-k},\dots,x_{m_{\tilde\lambda}}t^{-k},q^{\tilde\lambda_{m_{\tilde\lambda}+1}}t^{N-m_{\tilde\lambda}-k-1},
	% \dots,q^{\tilde\lambda_{N-k-1}}t^{-k+1},q^{\tilde\lambda_{N-k}}t^{-k})\mathop=^{(*)}\\\displaystyle\prod_{j=0}^{\max\tilde \lambda-1}
	% \prod_{i=1}^{m_{\tilde\lambda}}(x_{i}-q^{j}t^{k+\lfloor\tilde \lambda\rfloor_{j}}).\end{array}
	% \]
	% Since $\max\tilde\lambda=\max\lambda$, $m_{\lambda}=m_{\tilde\lambda}$, and $\lfloor\lambda\rfloor_{j}=k+\lfloor\tilde\lambda\rfloor_{j}$, we find our result.
 \end{enumerate}\qed
 
 \begin{example}\rm
	 Consider $\lambda=[664331110]$, we alternatively use Propositions \ref{saturate} and \ref{standard} for computing 
	 $MS_{664331110}(\mathbb X_{2}+\lbag4331110\rbag;q,t)$:
	 \[{}
	 \begin{array}{c}
		MS_{664331110}(\mathbb X_{2}+\lbag4331110\rbag;q,t)\\
		 \updownarrow \mbox{Prop. \ref{standard}}\\
		 MS_{66433111}(t^{-1}\mathbb X_{2}+\lbag433111\rbag;q,t)\\
		 \updownarrow \mbox{Prop. \ref{saturate}}\\
		 (t^{-1}x_{1}-1)(t^{-1}x_{2}-1)MS_{55322000}(t^{-1}q^{-1}\mathbb X_{2}+\lbag322000\rbag;q,t)\\
		 \updownarrow \mbox{Prop. \ref{standard}}\\
		 (t^{-1}x_{1}-1)(t^{-1}x_{2}-1)MS_{55322}(t^{-4}q^{-1}\mathbb X_{2}+\lbag322\rbag;q,t)\\
		 \updownarrow \mbox{Prop. \ref{saturate}}\\
		 \displaystyle\prod_{i=1}^{2}(t^{-1}x_{i}-1)(t^{-4}q^{-1}x_{i}-1)
		 (t^{-4}q^{-1}x_{i}-q)MS_{33100}(t^{-4}q^{-3}\mathbb X_{2}+\lbag100\rbag)\\
		 \updownarrow \mbox{Prop. \ref{standard}}\\
		 \displaystyle\prod_{i=1}^{2}(t^{-1}x_{i}-1)(t^{-4}q^{-1}x_{i}-1)
		 (t^{-4}q^{-1}x_{i}-q)MS_{331}(t^{-6}q^{-3}\mathbb X_{2}+\lbag1\rbag)\\
		 	 \updownarrow \mbox{Prop. \ref{saturate}}\\
		 	 \displaystyle\prod_{i=1}^{2}(t^{-1}x_{i}-1)(t^{-4}q^{-1}x_{i}-1)
		 (t^{-4}q^{-1}x_{i}-q)(t^{-6}q^{-3}x_{i}-1)MS_{220}(t^{-6}q^{-4}\mathbb X_{2}+\lbag0\rbag;q,t)\\
		  \updownarrow \mbox{Prop. \ref{standard}}\\
		   \displaystyle\prod_{i=1}^{2}(t^{-1}x_{i}-1)(t^{-4}q^{-1}x_{i}-1)
		 (t^{-4}q^{-1}x_{i}-q)(t^{-6}q^{-3}x_{i}-1)MS_{22}(t^{-7}q^{-4}\mathbb X_{2})\\
		  \updownarrow \mbox{Prop. \ref{saturate}}\\
		  \displaystyle\prod_{i=1}^{2}(t^{-1}x_{i}-1)(t^{-4}q^{-1}x_{i}-1)
		 (t^{-4}q^{-1}x_{i}-q)(t^{-6}q^{-3}x_{i}-1)(t^{-7}q^{-4}x_{i}-1)(t^{-7}q^{-4}x_{i}-q).
	 \end{array}
	 \]
	 As expected, the last polynomial is proportional to
	 \[{}
	 \prod_{i=1}^{2}(x_{i}-t)(x_{i}-qt^{4})
		 (x_{i}-q^{2}t^{4})(x_{i}-q^{3}t^{6})(x_{i}-q^{4}t^{7})(x_{i}-q^{5}t^{7}).
	\]
 \end{example}
 \section{Specializations of the type $t^{\alpha}q^{\beta}=1$ and quasi-staircase partitions \label{Spec}}
 %Set $\{\lambda\}=\sum \langle \lambda\rangle[i]$.
%First we recall that for some specialization of $(q,t)$ the quasi-staircase polynomials are singular (results contained in [JL]).
%Using an argument of dimension of eigenspaces, we show that the shifted Macdonald are homogeneous.\\ \\
\subsection{Admissible specialization}
We  denote by $QS(\ell,k;s,r;\beta)$ the \emph{quasi-staircase partition}
\begin{eqnarray*}
\left[((\beta+1)s+r)^{k},(\beta s+r)^{\ell},\dots,(s+r)^{\ell},
0^{r{l+1\over s-1}+\beta\ell}\right],
\end{eqnarray*}
where $r{\ell+1\over s-1}$ is an integer.
 We consider also, as in \cite{FM}, a specialization of $(t,q)=\left(u^{s-1\over g},\omega_{1}u^{-{\ell+1\over g}}\right)$ 
 where $g=\gcd(\ell+1,s-1)$ and $\omega_{1}^{s-1\over g}$ is a primitive $g$th root of the unity. 
 We call this kind of specialization a $(s,\ell)$-\emph{admissible specialization}.
 \begin{example}
	 \rm Let us illustrate the notion of $(s,\ell)$-admissible specialization by giving a few examples and counter-examples.
	 \begin{itemize}
		 \item $(t,q)=(u,u^{-3})$ is $(2,2)$-admissible.
		 \item $(t,q)=(u^{2},u^{-5})$ and $(t,q)=(u^{2},-u^{-5})$ are $(3,4)$-admissible.
		 \item $(t,q)=(u,-u^{-2})$ is $(3,3)$-admissible while $(t,q)=(u,u^{-2})$ is not $(3,3)$-admissible.
		 \item $(t,q)=(u,e^{2i\pi\over 3}u^{-2})$ and $(t,q)=(u,e^{2i\pi\over 3}u^{-2})$ are $(4,5)$-admissible but $(t,q)=(u,u^{-2})$ is not.
		 \item $(t,q)=(u,iu^{-2})$ and $(t,q)=(u,-iu^{-2})$ are $(5,7)$-admissible, while $(t,q)=(u,u^{-2})$ and $(t,q)=(u,-u^{-2})$ are not $(5,7)$-admissible.
	 \end{itemize}
 \end{example}
 Notice that the reason for the
definition of $\omega_{1}$ is given in the following:
\begin{lemma}\label{t^{a}q^{b}}
Suppose $t^{\alpha}q^{\gamma}=1$, $t=u^{b}$, and $q=\omega_{1}u^{-a}$. Then,
$\alpha=p\left(  \ell+1\right)  $ and $\gamma=p\left(  s-1\right)  $, for some
$p\in\mathbb{N}$.
\end{lemma}
{\it Proof.}
By hypothesis $\omega_{1}^{\gamma}u^{\alpha b-\gamma a}=1$. From $\gcd\left(
a,b\right)  =1$, it follows that $\alpha=ca$ and $\gamma=cb$, with
$c\in\mathbb{N}$. Thus $\left(  \omega_{1}^{b}\right)  ^{c}=1$ and $c=pg$, for
some $p\in\mathbb{N}$, because $\omega_{1}^{b}$ is a primitive $g$th root of
unity. Hence $\alpha=pga=p\left(  \ell+1\right)  $ and $\gamma=pgb=p\left(
s-1\right)  $.\qed

 \subsection{On the reciprocal sum $\lbag QS(\ell,k;s,r;\beta)\rbag$}
 In this section we prove that the intersection of the eigenspace of $\xi$, with eigenvalue $\lbag QS(\ell,k;s,r;\beta)\rbag_{q^{-1},t^{-1}}$, and the space
 generated by $P_{\mu}$ with $\mu\subseteq QS(\ell,k;s,r;\beta)$, has dimension $1$.\\
 We need the following technical lemma.
 \begin{lemma}\label{subseteq}
	Let $1\leq k\leq\ell$, $2\leq s$, $0\leq r,\beta$ and $0\leq\alpha\leq r{\ell+1\over s-1}$ be six integers and $\lambda= \left[(\beta s+r)^{k},((\beta-1) s+r)^{\ell},\dots,(s+r)^{\ell},0^{\ell+\alpha}\right]$.
		Let $\mu\subseteq\lambda$ such that there exists $i$ satisfying 
		$\langle\mu\rangle[i]=\langle\lambda\rangle[i]$. Then $i=k$ and 
		$\mu[1]=\lambda[1],\dots,\mu[k]=\lambda[k]$.
	\end{lemma}
	{\it Proof. } We have
	\begin{equation*}
\omega_{1}^{\mu[i]}u^{\frac1g((N-i)(s-1)-\mu[i](\ell+1))}=\omega_{1}^{\beta}u^{\frac1g((s-1)(N-k)-(\beta s+r)(\ell+1)}.
\end{equation*} 
In other words, this means that $(i,\mu[i])$ lies on the line 
\begin{equation}\label{linemu}
y=\beta s+r+{s-1\over\ell +1}(k-x).
\end{equation}
%\[{}
%k-i=\left(\mu[i]-(\beta s+r))\right){\ell+1\over s-1}.
%\]
Since $\mu\subset\lambda$, one has $\mu[i]\leq\beta s+r$ and so, $k\leq i$.
We assume first that $\lambda[i]>0$ (i.e. $i\leq \beta\ell+k$). In this case $(i,\lambda[i])$ lies on or below the line
\begin{equation}\label{linelambda}
y=\beta s+r+{s\over \ell}(k-x).
\end{equation}
Hence, 
\begin{equation}\label{m-l}
0\geq \mu[i]-\lambda[i]\geq {s+\ell\over \ell(\ell+1)}(i-k).
\end{equation}
In other words, $i\leq k$ and so, $i=k$.\\
Now, suppose that $i>k+(\beta-1)\ell$. In this case  we have $\mu[i]=0$ and, from \eqref{linemu}, we obtain
\begin{equation*}
i=k+(\beta s+r){\ell+1\over s-1}=r{\ell+1\over s-1}+\beta+{s\over s-1}(\ell+1)+k>{}
r{\ell+1\over s-1}+\beta\ell+k.
\end{equation*} 
Then, $i$ is strictly greater than the size of $\mu$. But this is impossible.
Hence, the only remaining possibility is $k=i$ and so $\mu[k]=\lambda[k]$. Since $\mu\subset \lambda$, we deduce also 
$\mu[1]=\cdots\mu[k]=\lambda[k]$. \qed
 %satisfying $t^{\ell+1}q^{s-1}=1$ and $t^{a}q^{b}\neq 1$ for $1\leq a,b\leq g$  for %as in \cite{JL} the specialization
 %$g=\gcd(\ell+1,s-1\}$. %Under the considered specialization we have $q^{s-1\over g}=\exp\left\{2i\pi d\over g\right\}t^{-{\ell+1\over g}}$ for some $d\in \{0,\dots,g-1\}$. 
%Hence,
%\[{}
%q=\exp\left\{2i\pi(d+gd')\over s-1\right\}t^{-{\ell+1\over s-1}},
%\]
%with $0\leq d'<{s-1\over g}$. Let $a, b\in\mathbb N$ such that
%\[q^{a}t^{b}=\exp\left\{2i\pi(d+gd')a\over s-1\right\}t^{b-a{\ell+1\over s-1}}.\]
%We show now a necessary condition for $t^{a}q^{b}\neq 1$ when 
%$1\leq a,b \leq g$. The properties $t^{a}q^{b}\neq 1$  implies $b=a{\ell+1\over s-1}$ that is ${s-1\over g}$ divides $a$. 
%In this case $\exp\left\{2i\pi(d+gd')a\over s-1\right\}\neq 1$ and this means that $s-1$ does not divide $(d+d'g)a$. 
%Since ${s-1\over g}$ divides $a$,
%$s-1$ divides $d'ga$ and necessarily $da$ is not a multiple of $s-1$. In other words, since   ${s-1\over g}$ divides $a$, $g$ does not divide $d$ and then $d\neq 0$.
%Let $u$ such that $t=u^{s-1\over g}$ and $u^{-{\ell+1\over g}}=\exp\left\{2i\pi(d-1+d'g)\over s-1\right\}t^{-{\ell+1\over s-1}}$.
%In this case, one obtains $q=u^{-{\ell+1\over g}}\exp\left\{2i\pi(1+d'g)\over s-1\right\}$.

%\[{}
%(t,q)=\left(u^{s-1\over g},u^{-{\ell+1\over g}}\omega_{1}\right)
%\] 
%satisfying $g=\gcd(\ell+1,s-1)$ and $\omega_{1}=\exp\left\{2i(1+dg\over s-1\right\}$ with $\omega_{1}^{r}=1$.
\begin{example}
	\rm{}
	In Figure \ref{Lines},  we illustrate the proof of Lemma \ref{subseteq} for the parameters $s=4$, $\ell=4$, $k=3$, $\beta=5$ and $r=3$.
	The line defined by  (\ref{linemu}) is drawn in black, the line defined by (\ref{linelambda}) in green, and the partition $\lambda$ in blue. 
	The area filled in red illustrates the inequality (\ref{m-l}).
	\begin{figure}[h]
	\begin{center}
		\includegraphics[width=7cm]{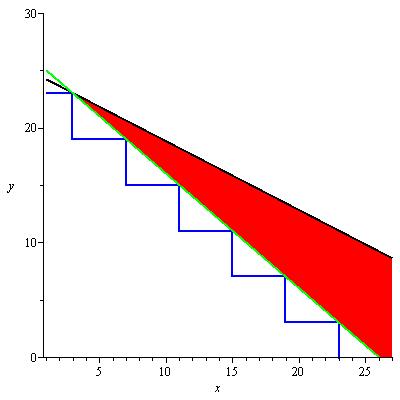}
		\end{center}
		\caption{Illustration of the proof of Lemma. \ref{subseteq} \label{Lines}}
	\end{figure}

%We consider the specialization $(t,q)=\left(u,u^{-3}\right)$ which is $(2,2)$-admissible.
%Let $\mu\subset \lambda=QS(4,2;2,0;2)=[6,6,4,4,2,2,0,0]$ such that there exists $i$ with $\langle\mu\rangle[i]=\langle\lambda[2]\rangle=q^{6}t^{6}=u^{-12}$.
%Since $\langle\mu\rangle=u^{8-i-3\mu[i]}$ we deduce that 
%$i=20-3\mu[i]$. But we have also $\mu[i]\leq 6$. Hence, $i\geq 2$. Remark that if $i>2$ then $\mu[i]\leq 4$ and so $i\geq 8$ but this implies $\mu[i]=0$ and so $i>20$.
%Since the size of the partition is $8$ the only remaining case is $i=2$ and $\mu[2]=6$. Now $6\geq \mu[1]\geq\mu[2]$ implies $\mu[1]=6$.
\end{example}
\begin{proposition}\label{eigenvalues}
If $\mu\subseteq QS(\ell,k;s,r;\beta)$ is such that $\lbag\mu\rbag=\lbag QS(\ell,k;s,r;\beta)\rbag$, then 
	$\mu=QS(\ell,k;s,r;\beta)$. 
\end{proposition}
{\it Proof.}
We set $\lambda=QS(\ell,k;s,r;\beta)$.
Let $\mu\subseteq \lambda$ is such that $\lbag\mu\rbag=\lbag\lambda\rbag$.{}
 We proceed by induction on $\beta+k$. 
The starting point of the induction is $\beta+k=0$ and this implies $\mu=QS(\ell,0;s,0;0)=[0^{N}]$.
There exists $i$ such that  $\langle \mu\rangle[i]=\langle \lambda\rangle[k]$.
We have to consider two cases:
\begin{itemize}
	\item
If $k>0$ then, from Lemma \ref{subseteq}, we find $i=k$ and $\mu[1]=\cdots=\mu[k]=(\beta+1) s+r$.
	We remark that $[QS(\ell,k;s;\beta)[k+1],\dots,QS(\ell,k;s;\beta)[N]]=QS(\ell,0;s,r;\beta)$ and
	$[\langle \mu\rangle[k+1],\cdots,\langle\mu\rangle[N]]=\langle[ \mu[k+1],\cdots,\mu[N]]\rangle$. In fact, $\lbag[ \mu[k+1],\cdots,\mu[N]]\rbag
	=\lbag QS(\ell,0;s,k;\beta)\rbag$ and $[ \mu[k+1],\cdots,\mu[N]]\subseteq QS(\ell,0;s,k;\beta)$.
	 Hence, we can use the induction hypothesis to obtain  
	$[\mu[k+1],\dots,\mu[N]]=QS(\ell,0;s,k;\beta)$, which implies our result.
\item
If $k=0$ then, from Lemma \ref{subseteq},  we find $i=\ell$ and $\mu[1]=\cdots=\mu[\ell]=\beta s$.
	We remark that $[QS(\ell,0;s,r;\beta)[\ell+1],\dots,QS(\ell,0;s,k;\beta)[N]]=QS(\ell,0;s,k;\beta-1)$  and
	$[\langle \mu\rangle[\ell+1],\cdots,\langle\mu\rangle[N]]=\langle[ \mu[\ell+1],\cdots,\mu[N]]\rangle.$
	So $\lbag[ \mu[\ell+1],\cdots,\mu[N]]\rbag
	=\lbag QS(\ell,0;s,k;\beta-1)\rbag$ and $[ \mu[\ell+1],\cdots,\mu[N]]\subseteq QS(\ell,0;s,k;\beta-1)$.
	 Hence, we can use the induction hypothesis to obtain  
	$[\mu[\ell+1],\dots,\mu[N]]=QS(\ell,0;s,k;\beta-1)$  and this implies our result.
	\end{itemize}
\qed
\begin{example}
		\rm The following table contains all the reciprocal sums $\lbag\mu \rbag$ associated to the partitions $\mu\subseteq[4,2,0]${}
		\[{}
		\begin{array}{|c|c|c|c|}
		\hline{}
		\lbag420\rbag&\lbag320\rbag&\lbag220\rbag&\lbag110\rbag\\
		q^{4}t^{2}+q^{2}t+1&q^{3}t^{2}+q^{2}t+1&q^{2}t^{2}+q^{2}t+1&qt^{2}+qt+1\\\hline
		\lbag410\rbag&\lbag310\rbag&\lbag210\rbag&\lbag100\rbag\\
		q^{4}t^{3}+qt+1&q^{3}t^{2}+qt+1&q^{2}t^{2}+qt+1&qt^{2}+t+1\\\hline
		\lbag400\rbag&\lbag300\rbag&\lbag200\rbag&\lbag000\rbag\\
		q^{4}t^{2}+t+1&q^{3}t^{2}+t+1&q^{2}t^{2}+t+1&t^{2}+t+1\\\hline
		\end{array}
		\]
		Under the $(2,2)$-admissible specialization $(t,q)=(u,u^{-2})$, the table becomes
		\[{}
		\begin{array}{|c|c|c|c|}
		\hline{}
		\lbag420\rbag&\lbag320\rbag&\lbag220\rbag&\lbag110\rbag\\
		u^{-6}+u^{-3}+1&u^{-4}+u^{-3}+1&u^{-3}+u^{-2}+1&u^{-1}+2\\\hline
		\lbag410\rbag&\lbag310\rbag&\lbag210\rbag&\lbag100\rbag\\
		u^{-6}+u^{-1}+1&u^{-4}+u^{-1}+1&u^{-2}+u^{-1}+1&2+u\\\hline
		\lbag400\rbag&\lbag300\rbag&\lbag200\rbag&\lbag000\rbag\\
		u^{-6}+1+u&u^{-4}+1+u&u^{-2}+1+u&1+u+u^{2}\\\hline
		\end{array}
		\]
		We observe that the only partition  whose reciprocal sum equals $u^{-6}+u^{-3}+1$ is $[420]$. 
	\end{example}
	Notice that Proposition \ref{eigenvalues} can be alternatively stated as follows.
	\begin{corollary}\label{ceigenvalues}
		Suppose that $\mu\subseteq \lambda=QS(\ell,k;s,r;\beta)$ is a partition
	such that $\langle\mu\rangle$ is a permutation of $\langle\lambda\rangle$. Then,
	\begin{itemize}
		\item
	$\mu=\lambda$,{}
	\item the intersection of the eigenspace of $\xi$ with eigenvalue $\lbag \lambda\rbag_{q^{-1},t^{-1}}$, and the space
 generated by $P_{\mu}$, with $\mu\subseteq \lambda$, has dimension $1$,
 \item the intersection of the eigenspace of $\Xi$ with eigenvalue $\lbag \lambda\rbag_{q^{-1},t^{-1}}$, and the space
 generated by $MS_{\mu}$, with $\mu\subseteq \lambda$, has dimension $1$.
 \end{itemize}
	\end{corollary}   
	{\it Proof.} It is easy to see that for a specialization of type $(t,q)=(u^{b},\omega_{1}u^{a})$ the following four assertions are equivalent:
	\begin{enumerate}
		\item  $\langle\mu\rangle$ is a permutation of $\langle\lambda\rangle$,
		\item  $\langle\mu\rangle_{q^{-1},t^{-1}}$ is a permutation of $\langle\lambda\rangle_{q^{-1},t^{-1}}$,
		\item  $\lbag\mu\rbag=\lbag\lambda\rbag$,{}
		\item  $\lbag\mu\rbag_{q^{-1},t^{-1}}=\lbag\lambda\rbag_{q^{-1},t^{-1}}$.
	\end{enumerate}
	Hence, Proposition \ref{eigenvalues} allows us to complete the proof.
	\qed
	\subsection{On the reciprocal vector $\langle QS(\ell,k;s,r;\beta)\rangle$}
	In this section, we prove that all the entries of $\langle QS(\ell,k;s,r;\beta)\rangle$ are distinct.
	Denote $QS\left(  \ell,k;s,r;\beta\right)  $ by $\lambda$.   Define the utility
function $h\left(  w,z\right)  =bz-aw$, so that, under the considered specialization, one has $\left\langle
\mu\right\rangle \left[  i\right]  =\omega_{1}^{\mu\left[  i\right]
}u^{h\left(  \mu\left[  i\right]  ,N-i\right)  }$. Define also%
\begin{equation*}
j_{\beta+2}=0;~j_{m}=k+\left(  \beta-m+1\right)  \ell,~1\leq m\leq\beta+1,
\end{equation*}
so that $j_{1}=N-n_{0}$, with $n_{0}=(\ell+1){r\over s-1}+\ell$. Suppose $j_{m+1}<i\leq j_{m}$. Then $\lambda\left[
i\right]  =r+ms$ and $\left\langle \lambda\right\rangle \left[  i\right]
=\omega_{1}^{r+ms}u^{h\left(  r+ms,N-i\right)  }$. If $j_{1}<i\leq N$ then
$\lambda\left[  i\right]  =0$ and $\left\langle \lambda\right\rangle
\left[  i\right]  =u^{h\left(  0,N-i\right)  }$.
\begin{proposition}
The entries of $\left\langle \lambda\right\rangle $ are pairwise distinct.
\end{proposition}
{\it Proof.}
If $j_{m+1}<i\leq j_{m}$ then $h\left(  r+ms,N-i\right)  =b\left(  N-i\right)
-a\left(  r+ms\right)  $.  Let $c\in\mathbb N$ such that $n_{0}=\left(  \ell+1\right)  c-1$ and $r=\left(  s-1\right)  \left(  c-1\right)
$. One has
\begin{multline}
h\left(  r+ms,N-j_{m}\right)     =-\frac{m\left(  s+\ell\right)  }{g}\leq
h\left(  r+ms,N-i\right)  
  \leq h\left(  r+ms,N-j_{m}+\ell-1\right) = \\ = \frac{1}{g}\left\{  -m\left(
s+\ell\right)  +\left(  s-1\right)  \left(  \ell-1\right)  \right\}.
\end{multline}
Suppose $1\leq i_{1}<i_{2}\leq N$. If $j_{m+1}%
<i_{1}<i_{2}\leq j_{m}$, then $h\left(  r+ms,N-i_{1}\right)  -h\left(
r+ms,N-i_{2}\right)  =b\left(  i_{2}-i_{1}\right)  >0$, or if $j_{1}%
<i_{1}<i_{2}\leq N$, then $h\left(  0,N-i_{1}\right)  -h\left(  0,N-i_{2}%
\right)  =b\left(  i_{2}-i_{1}\right)  >0$. Thus $\left\langle
\lambda\right\rangle \left[  i_{1}\right]  \neq \left\langle \lambda
\right\rangle \left[  i_{2}\right]  $. Now suppose $j_{m_{1}+1}<i_{1}\leq
j_{m_{1}}\leq j_{m_{2}+1}<i_{2}\leq j_{m_{2}}$, then the above bound shows%
\begin{equation*}
h\left(  r+m_{2}s,i_{2}\right)  -h\left(  r+m_{1}s,i_{1}\right)  \geq\frac
{1}{g}\left\{  \left(  m_{1}-m_{2}\right)  \left(  s+\ell\right)  -\left(
s-1\right)  \left(  \ell-1\right)  \right\}  .
\end{equation*}
Thus if $m_{1}-m_{2}>\dfrac{\left(  s-1\right)  \left(  \ell-1\right)
}{s+\ell}$, then $\left\langle \lambda\right\rangle \left[  i_{1}\right]
\neq \left\langle \lambda\right\rangle \left[  i_{2}\right]  $. Consider
the case $1\leq m_{1}-m_{2}\leq\dfrac{\left(  s-1\right)  \left(
\ell-1\right)  }{s+\ell}$; then $\left\langle \lambda\right\rangle \left[
i_{1}\right]  =\omega_{1}^{r+m_{1}s}u^{h\left(  r+m_{1}s,N-i_{1}\right)}  $ and
$\left\langle \lambda\right\rangle \left[  i_{2}\right]  =\omega
_{1}^{r+m_{2}s}u^{h\left(  r+m_{2}s,N-i_{2}\right)}  $. There are two different
arguments depending on whether $\omega_{1}\neq1$, equivalently $\gcd\left(
\ell+1,s-1\right)  >0$.

If $\omega_{1}\neq1$, suppose by way of contradiction that $1=\left\langle
\lambda\right\rangle \left[  i_{1}\right]  /\left\langle \lambda
\right\rangle \left[  i_{2}\right]  =q^{\left(  m_{1}-m_{2}\right)  s}%
t^{i_{2}-i_{1}}$. By  Lemma \ref{t^{a}q^{b}}, $\left(  m_{1}-m_{2}\right)  s=p\left(
s-1\right)  $, for some $p\in\mathbb{N}$, but $1\leq m_{1}-m_{2}\leq
\dfrac{\left(  s-1\right)  \left(  \ell-1\right)  }{s+\ell}<s-1$ and
$\gcd\left(  s,s-1\right)  =1$, which is a contradiction. Suppose
$j_{m+1}=j_{m}<i_{1}\leq j_{m}$ and $j_{1}<i_{2}\leq N$. By the above bound, and the fact that $h\left(  0,N-i_{2}\right)  \geq0$,
\begin{multline*}
h\left(  0,N-i_{2}\right)  -h\left(  r+m_{1}s,N-i_{1}\right)   
\geq-h\left(  r+m_{1}s,N-j_{m_{1}}+\ell-1\right)= \\
  =\frac{1}{g}\left\{  m_{1}\left(  s+\ell\right)  -\left(  s-1\right)
\left(  \ell-1\right)  \right\}  .
\end{multline*}
If $m_{1}>\frac{\left(  s-1\right)  \left(  \ell-1\right)  }{s+\ell}$, then
$h\left(  0,N-i_{2}\right)  -h\left(  r+m_{1}s,N-i_{1}\right)  >0$ and 
$\left\langle \lambda\right\rangle \left[  i_{1}\right]  \neq
\left\langle \lambda\right\rangle \left[  i_{2}\right]  $. Otherwise, 
suppose $1=\left\langle \lambda\right\rangle \left[  i_{1}\right]
/\left\langle \lambda\right\rangle \left[  i_{2}\right]  =q^{m_{1}%
s+r}t^{i_{2}-i_{1}}$ and $m_{1}s+r=p\left(  s-1\right)  $. Then, $m_{1}s=\left(
s-1\right)  \left(  p-c+1\right)  $, which is impossible for $1\leq m_{1}<s-1.$

Suppose $\gcd\left(  s-1,\ell+1\right)  =1$. Then $h\left(  \lambda\left[
i\right]  ,N-i\right)  \equiv-\left(  \ell+1\right)  \lambda\left[  i\right]
\operatorname{mod}\left(  s-1\right)  $. If $j_{m+1}<i\leq j_{m}$, then
$h\left(  \lambda\left[  i\right]  ,N-i\right)  \equiv m\left(  s+\ell\right)
\operatorname{mod}\left(  s-1\right)  $ and $\gcd\left(  s+\ell,s-1\right)
=1$. In the case $j_{m_{1}+1}<i_{1}\leq j_{m_{1}}\leq j_{m_{2}+1}<i_{2}\leq
j_{m_{2}}$, with $1\leq m_{1}-m_{2}\leq\dfrac{\left(  s-1\right)  \left(
\ell-1\right)  }{s+\ell}$ it follows that \begin{equation*}
h\left(  r+m_{2}s,i_{2}\right)
-h\left(  r+m_{1}s,i_{1}\right)  \equiv\left(  m_{1}-m_{2}\right)  \left(
s+\ell\right)  \operatorname{mod}\left(  s-1\right)  .\end{equation*} As before, this implies
$h\left(  r+m_{2}s,i_{2}\right)  -h\left(  r+m_{1}s,i_{1}\right)
\not\equiv 0\operatorname{mod}\left(  s-1\right)  $, since $1\leq m_{1}-m_{2}<s-1$.
The similar argument applies when $j_{1}<i_{2}\leq N$.

This concludes the proof.
\qed
{}
\begin{example}\rm
Here is an example showing that $h$ alone does not suffice to separate the
$\left\langle \lambda\right\rangle $ values. Let $\ell=5,s=3,N=15,n_{0}=5$. Then $g=2$ and $\lambda=\left[  6^{5},3^{5},0^{5}\right]  $ and the respective
values of $h$ are%
\[
\left[  -4,-5,-6,-7,-8,0,-1,-2,-3,-4,4,3,2,1,0\right] .
\]
However $\left\langle \lambda\right\rangle \left[  1\right]  =\omega
^{6}u^{-4}$ and $\left\langle \lambda\right\rangle \left[  10\right]
=\omega^{3}u^{-4}$, where $\omega=-1$. But also $\left\langle \lambda
\right\rangle \left[  6\right]  =\omega^{3}\neq \left\langle \lambda
\right\rangle \left[  15\right]  =1$.
\end{example}

 \section{Factorizations and wheel condition\label{wheel}}
In this section we investigate the case of the staircase partitions that are partitions of the form 
$$St(\ell,k;s;\beta):=[((\beta+1)s)^{k},(\beta s)^{\ell},\dots,s^{\ell},0^{\ell}]=QS(\ell,k;s,0;\beta),$$
for any $\ell\geq 1, 0\leq k<\ell, s\geq 2$ and $\beta\geq 1$. 
We also assume that the parameters $q$ and $t$ specialize as 
\begin{equation*}
(t,q)=\left(u^{s-1\over g},u^{-{\ell+1\over g}}\omega_{1}\right),
\end{equation*} 
where $g=\gcd(\ell+1,s-1)$ and $\omega_{1}^{s-1\over g}$ is a primitive $g$th root of the unity.
\subsection{Wheel condition and admissible partitions}
In this section we recall the main results of \cite{FM}.
A symmetric polynomial $P(x_{1},\dots, x_{N})$ satisfies the $(s,\ell)$-\emph{wheel condition} if 
$$\left\{{x_{2}\over x_{1}},\dots,{x_{\ell+1}\over x_{\ell}},{x_{1}\over x_{\ell+1}}\right\}\subset \{t,tq,\dots, tq^{s-1}\}$$ implies that
 $P(x_{1},\dots, x_{N})=0$. It is easy to check that the set of the symmetric polynomials satisfying the wheel condition is
  an ideal. This ideal is denoted by $J^{\ell,s}_{N}$ in \cite{FM}. A \emph{$(\ell,s,N)$-admissible partition} is a partition $\lambda=[\lambda_{1},\dots,\lambda_{N}]$
  satisfying $\lambda_{i}-\lambda_{i+\ell}\geq s$ for any $i=1,\dots,N-\ell$. 
  The following theorem summarizes two results of \cite{FM}.
  \begin{theorem}\label{FM}
	  \begin{itemize}
		  \item One has
		  \begin{equation*}
		  J_{N}^{\ell,s}=\mathrm{span}\{P_{\lambda}(\mathbb X_{N};q,t):\ \lambda \mbox{ is }(\ell,s,N)-\mbox{admissible}\}.
		  \end{equation*}
		  \item The space $J_{N}^{\ell,s}$ is stable under the action of $L_{q,t}^{+}$. 
	  \end{itemize}
  \end{theorem}
Remarking that $P_{St(\ell,k;s;\beta)}(\mathbb X_{(\beta+1)\ell+k};q,t)$ is a minimal degree polynomial belonging to $J_{(\beta+1)\ell+k}^{\ell,s}$, the second part of
Theorem \ref{FM} implies the following result. 
\begin{proposition}\label{P=Mstaircase}
	 \begin{equation*}
		P_{St(\ell,k;s;\beta)}(\mathbb X_{(\beta+1)\ell+k};q,t)\mathop=^{(*)}MS_{St(\ell,k;s;\beta)}(\mathbb X_{(\beta+1)\ell+k};q,t).\end{equation*}
\end{proposition}
{\it Proof.}
	Since $MS_{St(\ell,k;s;\beta)}(\mathbb X_{(\beta+1)\ell+k};q,t)$ is in the kernel of $L^{+}_{q,t}=(1-q)(\xi-\Xi)$,{}
	 the polynomials $P_{St(\ell,k;s;\beta)}(\mathbb X_{(\beta+1)\ell+k};q,t)$
	and $MS_{St(\ell,k;s;\beta)}(\mathbb X_{(\beta+1)\ell+k};q,t)$ are in the same eigenspace of $\xi$ with reciprocal sum
	\begin{equation*}
	\lbag St(\ell,k;s;\beta)\rbag=\sum_{i=1}^{k}q^{(\beta+1)s}t^{N-i}+\sum_{i=0}^{\beta}\sum_{j=1}^{\ell}q^{\ell s}t^{N-k-(\beta-i)\ell-j}.
	\end{equation*} 
	Recall that one has
	\begin{equation*}
	MS_{St(\ell,k;s;\beta)}\mathop=^{(*)}P_{St(\ell,k;s;\beta)}+\sum_{\mu\subset{}
	St(\ell,k;s;\beta)}c_\mu(q,t)\cdot P_{\mu},
	\end{equation*}
	for some coefficients $c_\mu(q,t)$. Consider the spaces generated by the polynomials $P_{\mu}$, with $\mu\subseteq St(\ell,k;s;\beta)$. This space splits into
	several eigenspaces, and each of them is associated with a reciprocal sum $\lbag\mu\rbag$, for $\mu\subseteq St(\ell,k;s;\beta)$. 
	From Corollary \ref{ceigenvalues},
	the subspace of the eigenspace associated with
	the reciprocal sum $\lbag St(\ell,k;s;\beta)\rbag$ generated by the polynomials $MS_{\mu}$, with $\mu\subseteq St(\ell,k;s;\beta)$, 
	has dimension $1$.
	 It follows that $P_{St(\ell,k;s;\beta)}(\mathbb X_{(\beta+1)\ell+k};q,t)$ and
	$MS_{St(\ell,k;s;\beta)}(\mathbb X_{(\beta+1)\ell+k};w,t)$ are proportional.
	\qed
	\begin{example}\rm
		For instance, observe that the shifted Macdonald polynomial
		\[{}\begin{array}{l}
		MS_{20}(x_{1},x_{2};q=\frac1u,t=u^{2})=\frac{\left( {\it x_2}\,{\it x_1}\,u+{{\it x_1}}^{2}u+{{\it x_2}}^{2}u-{\frac {
{\it x_2}\,{\it x_1}}{u}}+{\it x_1}\,{u}^{2}{\it x_2}-{{\it x_1}}^{2}-{{
\it x_2}}^{2}-{\it x_2}\,{\it x_1} \right) }{ 
 \left(1- {u}^{-2} \right)  \left( 1-{u}^{-1} \right) ^2}\end{array}
		\]
		is homogeneous. And so, it is equal to $P_{20}(x_{1},x_{2};q=\frac1u,u^{2})$ up to a multiplicative factor.
	\end{example}
	
\subsection{Factorizations}
Let $\beta>0$.
 From Proposition \ref{P=Mstaircase}, the polynomial $MS_{St(\ell,k;s;\beta)}$ is homogeneous for $(\ell,s)$-admissible specializations. Then, assuming
 that $(t,q)$ satisfies such a specialization, one has
 \begin{equation*}
 \begin{array}{l}
 MS_{St(\ell,k;s;\beta)}\left(\mathbb X_{k+\beta\ell}+{1-t^{\ell}\over 1-t}y;q,t\right)=
 y^{{\ell s\beta(\beta-1)\over 2}+ks(\beta+1)}MS_{St(\ell,k;s;\beta)}\left({\mathbb X_{k+\beta\ell}\over y}+{1-t^{\ell}\over 1-t};q,t\right).
 \end{array}
 \end{equation*}
 Applying Proposition \ref{standard} one obtains
 \begin{equation*}
 \begin{array}{r}
MS_{St(\ell,k;s;\beta)}\left(\mathbb X_{k+\beta\ell}+{1-t^{\ell}\over 1-t};q,t\right)\displaystyle\mathop=^{(*)}
%\prod_{j=0}^{s-1}\prod_{i=0}^{\beta\ell+k}(x_{i}-t^{\ell}q^{j})
\mathcal R\left(\mathbb X_{\beta\ell+k};t^{\ell}{1-q^{s}\over 1-q}\right)MS_{ST(\ell,k;s;\beta-1)}\left(\mathbb X_{k+\beta\ell};q,t\right).
\end{array}
 \end{equation*}
 Again Proposition \ref{P=Mstaircase} shows that the polynomial $MS_{ST(\ell,k;s;\beta-1)}\left(\mathbb X_{k+\beta\ell}\right)$ is homogeneous.
 \begin{proposition}\label{FirstFactWheel}
	 \begin{equation*}
 MS_{St(\ell,k;s;\beta)}\left(\mathbb X_{k+\beta\ell}+y{1-t^{\ell}\over 1-t};q,t\right)\displaystyle\mathop=^{(*)}\displaystyle
%\prod_{j=0}^{s-1}\prod_{i=0}^{\beta\ell+k}(x_{i}-t^{\ell}q^{j}y)
\mathcal R\left(\mathbb X_{\beta\ell+k};yt^{\ell}{1-q^{s}\over 1-q}\right)MS_{ST(\ell,k;s;\beta-1)}
\left(q^{-s}t^{-\ell}\mathbb X_{k+\beta\ell};q,t\right).
 \end{equation*}
 \end{proposition}
 \begin{example}\rm
	 We consider the partition $[42200]=ST(2,1;2;1)$ and the specialization $(t,q)=(u,u^{-3})$. We have
	 \[{}
	 {MS_{42200}(x_{1},x_{2},x_{3},yt,y)\over 
	 MS_{200}(q^{-2}t^{-2}x_{1},q^{-2}t^{-2}x_{2},q^{-2}t^{-2}x_{3})}\mathop=^{(*)}\prod_{i=1}^{3}(x_{i}-u^{2}y)
	 (x_{1}-u^{-1}y).
	 \]
 \end{example}
 Set $\mathbb Y_{i,\beta}=y_{i}+y_{i+1}+\cdots+y_{\beta}$, for any $0\leq i\leq \beta$, and  consider the polynomial which is symmetric in both alphabets $\mathbb X_{k}$ 
 and $\mathbb Y_{0,\beta}$:
 \begin{equation*}
\mathcal P_{\ell,k;s;\beta}(\mathbb X_{k};\mathbb Y_{0,\beta}):=
MS_{St(\ell,k;s;\beta)}\left(\mathbb X_{k}+{1-t^{\ell}\over1-t}\mathbb Y_{0,\beta};q,t\right).
 \end{equation*}
Proposition \ref{FirstFactWheel} implies
 \begin{equation*}
 \begin{array}{l}
 \mathcal P_{\ell,k;s;\beta}(\mathbb X_{k};\mathbb Y_{0,\beta})\displaystyle\mathop=^{(*)}\displaystyle
% \prod_{j=0}^{s-1}\left[\prod_{i=0}^{k}(x_{i}-t^{\ell}q^{j}y_{0})
 %\prod_{i=1}^{\beta} \prod_{a=0}^{\ell-1}(t^{a}y_{i}-t^{\ell}q^{j}y_{0})\right]
 \mathcal R\left(\mathbb X_{k}+{1-t^{\ell}\over1-t}\mathbb Y_{1,\beta};t^{\ell}{1-q^{s}\over 1-q}y_{0}\right)
 \mathcal P_{\ell,k;s;\beta-1}(q^{-s}t^{\ell}\mathbb X_{k};q^{-s}t^{\ell}\mathbb Y_{1,\beta}).\end{array}
 \end{equation*}
 Iterating, one finds
 \begin{equation*}{}
 \begin{array}{l}
 \mathcal P_{\ell,k;s;\beta}(\mathbb X_{k};\mathbb Y_{0,\beta})\displaystyle\mathop=^{(*)}\displaystyle
 %\prod_{\alpha=0}^{\beta-1}\prod_{j=0}^{s-1}\left[\prod_{i=0}^{k}(x_{i}-t^{\ell}q^{j}y_{\alpha})
 %\prod_{i=\alpha+1}^{\beta} \prod_{a=0}^{\ell-1}(t^{a}y_{i}-t^{\ell}q^{j}y_{\alpha})\right]
  \prod_{\alpha=0}^{\beta-1}\mathcal R\left(\mathbb X_{k}+{1-t^{\ell}\over 1-t}\mathbb Y_{\alpha+1,\beta};t^{\ell}{1-q^{s}\over 1-q}y_{\alpha}\right)
  \mathcal P_{\ell,k;s;0}(q^{-\beta s}t^{-\beta\ell}\mathbb X_{k};y_{\beta}q^{-\beta s}t^{-\beta\ell}),
 \end{array}
 \end{equation*}
 with
 \begin{equation*}
 \mathcal P_{\ell,k;s;0}(\mathbb X_{k};y_{\beta})=
 MS_{St(\ell,k;s;0)}\left(\mathbb X_{k}+y_{\beta}{1-t^{\ell}\over 1-t};q,t\right).
 \end{equation*}
Once again applying Proposition \ref{P=Mstaircase} and  Proposition \ref{standard}, one gets the following result. 
 \begin{theorem}\label{FactWheel}
	 \begin{equation*}
		 \begin{array}{l}
 \mathcal P_{\ell,k;s;\beta}(\mathbb X_{k};\mathbb Y_{0,\beta})\displaystyle\mathop=^{(*)}\prod_{\alpha=0}^{\beta}\mathcal R
 \left(\mathbb X_{k}+{1-t^{\ell}\over 1-t}\mathbb Y_{\alpha+1,\beta};t^{\ell}{1-q^{s}\over 1-q}y_{\alpha}\right)\\=
 \displaystyle
 \prod_{\alpha=0}^{\beta}\prod_{j=0}^{s-1}\left[\prod_{i=1}^{k}(x_{i}-t^{\ell}q^{j}y_{\alpha})
 \prod_{i=\alpha+1}^{\beta} \prod_{a=0}^{\ell-1}(t^{a}y_{i}-t^{\ell}q^{j}y_{\alpha})\right].\end{array}
	 \end{equation*}
 \end{theorem}
 \begin{example}
	 \rm{}
	 For $(t,q)=(u,u^{-3})$, we have
\begin{multline*}
	 MS_{42200}(x_{1},ty_{1},y_1,ty_{0},y_{0})\displaystyle\mathop=^{(*)}(x_{1}-u^{2}y_{0})(x_{1}-u^{-1}y_{0})
	 (x_{1}-u^{2}y_{1})(x_{1}-u^{-1}y_{1})\times\\\times(y_{1}-u^{2}y_{0})(y_{1}-u^{-1}y_{0})
	 (uy_{1}-u^{2}y_{0})(uy_{1}-u^{-1}y_{0}).
\end{multline*}
 \end{example}
\section{Beyond the wheel conditions \label{notwheel}}

First we recall that, for some specialization of $(q,t)$, the quasistaircase polynomials satisfy the highest weight condition (results contained in \cite{JL}).
Using an argument of dimension of eigenspaces, we show that the shifted Macdonald are homogeneous.
%More precisely, in all this section, we will denote by $QS(\ell,k;s,r;\beta)$ the quasi-staircase partition $[((\beta+1)s+r)^{k},(\beta s+r)^{\ell},\dots,(s+r)^{\ell},
%0^{r{l+1\over s-1}+\beta\ell}]$ where $r{\ell+1\over s-1}$ is an integer. 
We also continue to consider the specialization
\begin{equation*}
(t,q)=\left(u^{s-1\over g},u^{-{\ell+1\over g}}\omega_{1}\right),
\end{equation*} 
satisfying that $g=\gcd(\ell+1,s-1)$ and $\omega_{1}^{s-1\over g}$ is a $g$th primitive root of the unity.
Recall the following result of \cite{JL}.
\begin{theorem}\label{thjl}
Let $\beta, s, r, k, \ell\in \N$, with $k\leq \ell$. Consider the
partition $\lambda=[((\beta+1)s+r)^k,(\beta
 s+r)^\ell,\dots,(s+r)^\ell]$. The polynomial
 $P_\lambda(x_1+\dots+x_n;q,t)$ is in the kernel of $L^{+}_{q,t}$
 when
 $
n={\ell+1\over s-1}r+\ell(\beta+1)+k
 $
 is an integer and
 \begin{equation}\label{specqt}
  (t,q)=\left(u^{s-1\over g},u^{-{\ell+1\over g}}\omega_1\right),
 \end{equation}
 where $g=\gcd(\ell+1,s-1)$ and $\omega_{1}$ is a $r$th root of the unity such that $\omega_{1}^{s-1\over g}$ is a 
 $g$th primitive root of the unity.
 % $\omega_1=\exp\left\{2i\pi(1+dg)\over
 %s-1\right\}$ if $d$ denotes an integer such that $w_1^r=1$.\\
 %Remark that the condition ${l+1\over s-1}r\in\N$ implies that
 %${s-1\over g}$ divides $r$ and the condition $\omega_1^r=1$
 %implies $g$ divides $r$. Hence, in all the cases, $s-1$ divides
 %$r$.
\end{theorem}

Remarking that $L^{+}_{q,t}=(1-q)\sum D_{i}=(1-q)\left(\sum\xi_{i}-\sum\Xi_{i}\right)$, we find that the shifted symmetric
 polynomial $MS_{QS(\ell,k;s,r;\beta)}$ is an
eigenfunction of  $\sum\xi_{i}$ having the same eigenvalue as 
$P_{QS(\ell,k;s,r;\beta)}$. But from Corollary \ref{ceigenvalues} the corresponding eigenspace has dimension $1$. This proves the following result. 
\begin{corollary}\label{P=MS}
\begin{equation*}
P_{QS(\ell,k;s,r;\beta)}\mathop=^{(*)}MS_{QS(\ell,k;s,r;\beta)}.
\end{equation*}
Moreover, this implies that $MS_{QS(\ell,k;s,r;\beta)}$ is homogeneous.
\end{corollary}

%\subsection{Vanishing properties and factorizations}

By Corollary \ref{P=MS}, we have
\begin{multline}
MS_{QS(\ell,k;s,r;\beta)}\left(\displaystyle\mathbb X_{k+\beta\ell}+{1-t^{r{\ell+1\over s-1}+\ell}\over 1-t}y_{\beta}\right)=\\=
y_{\beta}^{\frac12(\beta+1)(\ell\beta+2k)s+r(k+l\beta)}MS_{QS(\ell,k;s,r;\beta)}
\left(\displaystyle{\mathbb X_{k+\beta \ell}\over y_{\beta}}+{1-t^{r{\ell+1\over s-1}+\ell}\over 1-t}\right).
\end{multline}
Applying Proposition \ref{standard}, we obtain
\begin{multline}
\displaystyle
MS_{QS(\ell,k;s,r;\beta)}
\left(\displaystyle\mathbb X_{k+\beta \ell}+{1-t^{r{\ell+1\over s-1}+\ell}\over 1-t}\right)\displaystyle\mathop=^{(*)} \\\displaystyle\mathop=^{(*)}{}
\displaystyle 
%\prod_{j=0}^{s+r-1}\prod_{i=1}^{k+\beta \ell}\left(x_{i}-t^{r{\ell+1\over s-1}+\ell}q^{j}\right){}
\mathcal R\left(\mathbb X_{k+\beta\ell};t^{{\ell+1\over s-1}r+\ell}{1-q^{s+r}\over 1-q}\right)
MS_{QS(\ell,k;s,0;\beta)}\left(q^{-r-s}t^{-r{\ell+1\over s-1}-\ell}\mathbb X_{k+\beta\ell}\right).
\end{multline}
Again, by Corollary \ref{P=MS},  $MS_{QS(\ell,k;s,0;\beta)}$ is an homogeneous polynomial, and then,

\begin{multline}\label{MSQS}
\displaystyle
MS_{QS(\ell,k;s,r;\beta)}
\left(\displaystyle\mathbb X_{k+\beta\ell}+{1-t^{r{\ell+1\over s-1}+\ell}\over 1-t}y_{\beta}\right)\displaystyle\mathop=^{(*)} \\\displaystyle\mathop=^{(*)}{}
\displaystyle 
%\prod_{j=0}^{s+r-1}\prod_{i=1}^{k+\beta \ell}\left(x_{i}-t^{r{\ell+1\over s-1}+\ell}q^{j}y_{\beta}\right){}
\mathcal R\left(\mathbb X_{k+\beta\ell};t^{{\ell+1\over s-1}r+\ell}{1-q^{s+r}\over 1-q}y_{\beta}\right)
MS_{QS(\ell,k;s,0;\beta-1)}\left(q^{-r-s}t^{-r{\ell+1\over s-1}-\ell}\mathbb X_{k+\beta\ell}\right).
\end{multline}
We summarize this result in the following theorem.
\begin{theorem}\label{LastFactTheo}
	%\[\begin{array}{l}MS_{QS(\ell,k;s,r;\beta)}\left(\mathbb X_{k}+{1-t^{\ell}\over1-t}
	%\mathbb Y_{0,\beta-1}+{1-t^{r{\ell+1\over s-1}+\ell}\over 1-t}y_{\beta};q,t\right)\displaystyle\mathop=^{(*)}\\
	%\displaystyle 
%\prod_{j=0}^{s+r-1}\prod_{i=1}^{k+}\left(x_{i}-t^{r{\ell+1\over s-1}+\ell}q^{j}y_{\beta}\right){}
%%\prod_{\alpha=0}^{\beta}\prod_{j=0}^{s-1}\left[\prod_{i=0}^{k}(x_{i}-t^{\ell}q^{j}y_{\alpha})
 %\prod_{i=\alpha+1}^{\beta} \prod_{a=0}^{\ell-1}(t^{a}y_{i}-t^{\ell}q^{j}y_{\alpha})\right]
	%\end{array}\]
	%{\tt TODO: give the expression for 
	\begin{multline*}
	MS_{QS(\ell,k;s,r;\beta)}\left(\mathbb X_{k}+{1-t^{\ell}\over1-t}
	\mathbb Y_{0,\beta-1}+{1-t^{r{\ell+1\over s-1}+\ell}\over 1-t}y_{\beta};q,t\right) \displaystyle\mathop=^{(*)} \\
	\displaystyle\mathop=^{(*)}
	\mathcal R\left(\mathbb X_{k}+{1-t^{\ell}\over1-t}
	\mathbb Y_{0,\beta-1};t^{{\ell+1\over s-1}r+\ell}{1-q^{s+r}\over 1-q}y_{\beta}\right){}\\\displaystyle\times
	\prod_{\alpha=0}^{\beta-1}\mathcal R\left(\mathbb X_{k}+{1-t^{s}\over 1-t}\mathbb Y_{\alpha+1,\beta-1};t^{\ell}{1-q^{s}\over 1-q}y_{\alpha}\right)
	%\\
	%=
	%\displaystyle\prod_{j=0}^{s+r-1}\left[\prod_{i=1}^{k}\left(x_{i}-t^{r{\ell+1\over s-1}+\ell}q^{j}y_{\beta}\right){}
	%\prod_{a=0}^{\ell-1}\prod_{\alpha=0}^{\beta-1}(q^{a}y_{\alpha}-t^{r{\ell+1\over s-1}+\ell}q^{j}y_{\beta})
	%\right]\\\displaystyle
	%\times{}
	%\prod_{\alpha=0}^{\beta-1}\prod_{j=0}^{s-1}\left[\prod_{i=0}^{k}(x_{i}-t^{\ell}q^{j}y_{\alpha})
	%\prod_{i=\alpha+1}^{\beta-1}\prod_{a=0}^{\ell-1}(t^{a}y_{i}-t^{\ell}q^{j}y_{\alpha})\right].
	\end{multline*}
	%}
\end{theorem}
{\it Proof.} 
From equality (\ref{MSQS}), we obtain
\begin{multline*}
MS_{QS(\ell,k;s,r;\beta)}\left(\mathbb X_{k}+{1-t^{\ell}\over1-t}
	\mathbb Y_{0,\beta-1}+{1-t^{r{\ell+1\over s-1}+\ell}\over 1-t}y_{\beta};q,t\right)\displaystyle\mathop=^{(*)}
	\\\displaystyle\mathop=^{(*)}
	\displaystyle%\prod_{j=0}^{s+r-1}\left[\prod_{i=1}^{k}\left(x_{i}-t^{r{\ell+1\over s-1}+\ell}q^{j}y_{\beta}\right){}
	\mathcal R\left(\mathbb X_{k}+{1-t^{\ell}\over1-t}
	\mathbb Y_{0,\beta-1};t^{{\ell+1\over s-1}r+\ell}{1-q^{s+r}\over 1-q}y_{\beta}\right)
	%\prod_{a=0}^{\ell-1}\prod_{\alpha=0}^{\beta-1}(q^{a}y_{\alpha}-t^{r{\ell+1\over s-1}+\ell}q^{j}y_{\beta})
	%\right]
	\\\times
	MS_{QS(\ell,k;s,0;\beta-1)}\left(q^{-r-s}t^{-r{\ell+1\over s-1}-\ell}\left(\mathbb X_{k}+{1-t^{\ell}\over1-t}\mathbb Y_{0,\beta-1}\right);q,t\right).
\end{multline*}
The use of Theorem \ref{FactWheel} completes the proof.
\qed
\begin{example} For $(t,q)=(u,u^{-3})$, we have
	\[{}\begin{array}{l}
	MS_{533}(x_{1}+(1+t)y_{0}+(1+t+t^{2}+t^{3}+t^{4})y_{1};q=u^{-3},t=u)\\\displaystyle\mathop=^{(*)}
\mathcal R(x_{1}+(1+t)y_{0};t^{5}(1+q+q^{2})y_{1})
MS_{2}(x_{1}+(1+t)y_{0};q=u^{-3},t=u)\\\displaystyle\mathop=^{(*)}
\mathcal R(x_{1}+(1+t)y_{0};t^{5}(1+q+q^{2})y_{1})
\mathcal R(x_{1};t^{2}(1+q)y_{0})\\
=\displaystyle\prod_{i=0}^{2}\left[(x_{1}-t^{5}q^{i}y_{1})(y_{0}-t^{5}q^{i}y_{1})
(ty_{0}-t^{5}q^{i}y_{1})\right](x_{1}-t^{2}y_{0})(x_{1}-t^{2}qy_{0})\\
=(x_{0}-u^{5}y_{1})(x_{0}-u^{2}y_{1})(x_{0}-u^{-1}y_{1})(y_{0}-u^{5}y_{1})(y_{0}-u^{2}y_{1})(y_{0}-u^{-1}y_{1})\\\times
(uy_{0}-u^{5}y_{1})(uy_{0}-u^{2}y_{1})(uy_{0}-u^{-1}y_{1})(x_{1}-u^{2}y_{0})(x_{1}-u^{-1}y_{0}).
	\end{array}
	\]
\end{example}
%\begin{example}
%	\[{}\begin{array}{l}
%	MS_{533}(x_{1}+(1+t)y_{0}+(1+t+t^{2}+t^{3}+t^{4})y_{1};q=u^{-3},t=u)\displaystyle\mathop=^{(*)}\\\displaystyle
%	\prod_{i=0}^{1}(x_{1}-t^{2}q^{i}y_{0})\prod_{j=0}^{2}(x_{1}-t^{5}q^{j}y_{1})(y_{0}-t^{5}q^{j}y_{1})(qy_{0}-t^{5}q^{j}y_{1})=\\
%	\displaystyle{}
%	\prod_{i=0}^{1}(x_{1}-u^{2-3i}y_{0})\prod_{j=0}^{2}(x_{1}-u^{5-3j}y_{1})(y_{0}-u^{5-3j}y_{1})(y_{0}-u^{4-3j}y_{1})
%	\end{array}
%	\]
%\end{example}

\section{Conclusion and perspectives \label{Concl}}
\subsection{Second clustering property}
All the equations obtained in the paper specialize to Jack polynomials by sending $u$ to $1$ when $\ell+1$ and $s-1$ are coprime. This
implies also that $s-1$ divides $r$.
In that context Proposition \ref{FirstFactWheel} gives that
\begin{equation}\label{SCPeq1}
	J_{St(\ell,k;s;\beta)}^{\left(-{\ell+1\over s-1}\right)}\left(\mathbb X_{k+\beta\ell}+\ell y\right)\mathop=^{(*)}{}
	\prod_{i=1}^{\beta\ell}(x_{i}-y)^{s}J_{St(\ell,k;s;\beta-1)}(\mathbb X_{k+\beta\ell}).
\end{equation}
By the equality (\ref{MSQS}),
\begin{equation}\label{SCPeq2}
	J_{QS(\ell,k;s,r;\beta)}^{\left(-{\ell+1\over s-1}\right)}\left(\mathbb X_{k+\beta\ell}+\left(r{\ell+1\over s-1}+\ell\right) y\right){}
	\mathop=^{(*)}{}
	\prod_{i=1}^{\beta\ell}(x_{i}-y)^{s+r}J_{St(\ell,k;s;\beta-1)}(\mathbb X_{k+\beta\ell}).
\end{equation}
The fact that $\displaystyle\prod_{i=1}^{\beta\ell}(x_{i}-y)^{s+r}$ divides  the polynomial 
$J_{QS(\ell,k;s,r;\beta)}^{\left(-{\ell+1\over s-1}\right)}\left(\mathbb X_{k+\beta\ell}+\left(r{\ell+1\over s-1}+\ell\right) y\right)${}
is a special case of a result of  \cite{BGS} (Theorem 1.1).
Using the following table:
\[{}
\begin{array}{|c|c|}
	\hline \mbox{Our notation}&\mbox{Haldane notation}\\\hline
	\ell&k\\
	k&0\\
	s&r\\
	{r\over s-1}+1&s\\
	s+r&s(r-1)+1\\
	St(\ell,0;s;\beta)&\lambda^{\beta}_{k,r,1}\\
	QS(\ell,0;s,r;\beta)&\lambda^{\beta}_{k,r,s}\\
	r{\ell+1\over s-1}+\ell&n_{0}\\
	\mathbb X_{k+\beta\ell}&z_{n_{0}+1}+\cdots+z_{N}\\
	y&Z=z_{1}=\cdots=z_{n_{0}}\\\hline
\end{array}
\]
we recover the equality (\ref{SCC}) by setting $k=0$ in (\ref{SCPeq2}). This proves the second clustering property conjectured
by Bernevig and Haldane \cite{BH2}.

The formula (\ref{MSQS}) is more general than those conjectured in \cite{BH2} for two reasons. First, we consider quasistaircase partitions 
$QS(\ell,k;s,r;\beta)$, with $0\leq k< \ell$ (in \cite{BH1,BH2} only the case $k=0$ was investigated).
It should be interesting to know if some of these polynomials can be interpreted as wave functions in FQHT. Also, 
when $\omega_{1}\neq 1$,
the Macdonald polynomial does not degenerate to a Jack when $u$ tends to $1$.

More generally, the following equality is obtained from Theorem \ref{LastFactTheo}:
\begin{multline*}
J_{QS(\ell,k;s,r;\beta)}^{\left(-{\ell+1\over s-1}\right)}\left(\mathbb X_{k}+\ell\mathbb Y_{0,\beta-1}+
\left(r{\ell+1\over s-1}+\ell\right)y_{\beta}\right)
\displaystyle\mathop=^{(*)} \\\displaystyle\mathop=^{(*)}
\prod_{i=1}^{k}(x_{i}-y_{\beta})^{r+s}
\prod_{\alpha=0}^{\beta}\left[(y_{\alpha}-y_{\beta})^{(s+r)\ell}\prod_{i=1}^{k}(x_{i}-y_{\alpha})^{s}\prod_{i=\alpha+1}^{\beta-1}(y_{i}-y_{\alpha})^{\ell s}\right]
\end{multline*}
More general identities involves partitions which are not quasistaircase for the considered specialization. For instance,
\[{}
J_{43210}^{(-3)}(x+2y_{1}+2y_{2})\displaystyle\mathop=^{(*)}y_{1}y_{2}(y_{1}-y_{2})^{4}(x-y_{1})^{2}(x-y_{2})^{2}.
\]
This formula is in fact a specialization of
\[{}\begin{array}{rl}
P_{43210}\left(x+(1+t)y_{1}+(1+t)y_{2};q=t^{-3},t\right){}
\displaystyle\mathop=^{(*)}&y_{1}y_{2}(y_{1}-ty_{2})(y_{1}-t^{2}y_{2})
(y_{2}-ty_{1})(y_{2}-t^{2}y_{1})\\&(tx-y_{1})(x-t^{2}y_{1})(tx-y_{2})(x-t^{2}y_{2}).\end{array}
\]
Notice that this polynomial does not satisfy the highest weight condition. This suggests that there exists a Macdonald version of the result of \cite{BGS}, Theorem 1.1.  
A precise statement remains to be formulated.
\subsection{Other clustering/factorizations properties\label{other}}
The first and third clustering conjecture suggest that there exist many ways to factorize highest weight Macdonald polynomials by specializing the 
variables $x_{1},x_{2},\dots,x_{N}$. Let us illustrate this remark by giving an example. 

\begin{figure}[h]
	\begin{center}
\begin{tikzpicture}%
\GraphInit[vstyle=Shade]
    \tikzstyle{VertexStyle}=[shape = rectangle,
                             draw
]

\Vertex[x=0, y=0,
 L={$MS_{420}(\mathbb X_{111})$},style={shape=circle}%shading=ball,ball color = yellow!60}
]{111}
%\Vertex[x=9, y=6, L={$[201]$}]{z4}
\SetUpEdge[lw = 1.5pt,
color = orange,
 labelcolor = gray!30,
 labelstyle = {draw},
 style={post}
]
\Vertex[x=0, y=2,
 L={$MS_{530^{3}}(\mathbb X_{311})$},style={shape=circle}%shading=ball,ball color = yellow!60}
]{311}

\Vertex[x=-3, y=4,
 L={$MS_{640^{5}}(\mathbb X_{511})$},style={shape=circle}%shading=ball,ball color = yellow!60}
]{511}

\Vertex[x=3, y=4,
 L={$MS_{640^{5}}(\mathbb X_{331})$},style={shape=circle}%shading=ball,ball color = yellow!60}
]{331}

\Vertex[x=-6, y=6,
 L={$MS_{750^{7}}(\mathbb X_{711})$},style={shape=circle}%shading=ball,ball color = yellow!60}
]{711}

\Vertex[x=0, y=6,
 L={$MS_{750^{7}}(\mathbb X_{531})$},style={shape=circle}%shading=ball,ball color = yellow!60}
]{531}

\Vertex[x=6, y=6,
 L={$MS_{750^{7}}(\mathbb X_{333})$},style={shape=circle}%shading=ball,ball color = yellow!60}
]{333}

\Edge[label={$\times R_{1,3}$}](111)(311)
\Edge[label={$\times R_{1,5}$}](311)(511)
 \Edge[label={$\times R_{2,3}$}](311)(331)
 \Edge[label={$\times R_{1,7}$}](511)(711)
 \Edge[label={$\times R_{2,3}$}](511)(531)
 \Edge[label={$\times R_{1,5}$}](331)(531)
 \Edge[label={$\times R_{3,3}$}](331)(333)

%\tikzset{LabelStyle/.style = {draw,
 %                                    fill = yellow,
 %                                    text = red}}

%\GraphInit[vstyle=Shade]
 %   \tikzstyle{VertexStyle}=[shape = rectangle
%]
%\Vertex[x=0,y=6,L={$(i,i+1)$ if $v_{i}<v_{i+1}$},style={shape=circle}%shading=ball,ball color = yellow!60}
%]{bla}
%\Vertex[x=0,y=5,L={$v\cdot\Phi=[v_{2},\dots,v_{N},v_{1}+1]$},style={shape=circle}%shading=ball,ball color = yellow!60}
%]{bla}
\end{tikzpicture}
\end{center}
\caption{Computation of $MS_{4+r\ 2+r\ 0^{1+2r}}(\mathbb X_{\lambda};q=u^{-2},t=u).$\label{FigComp}}
\end{figure}
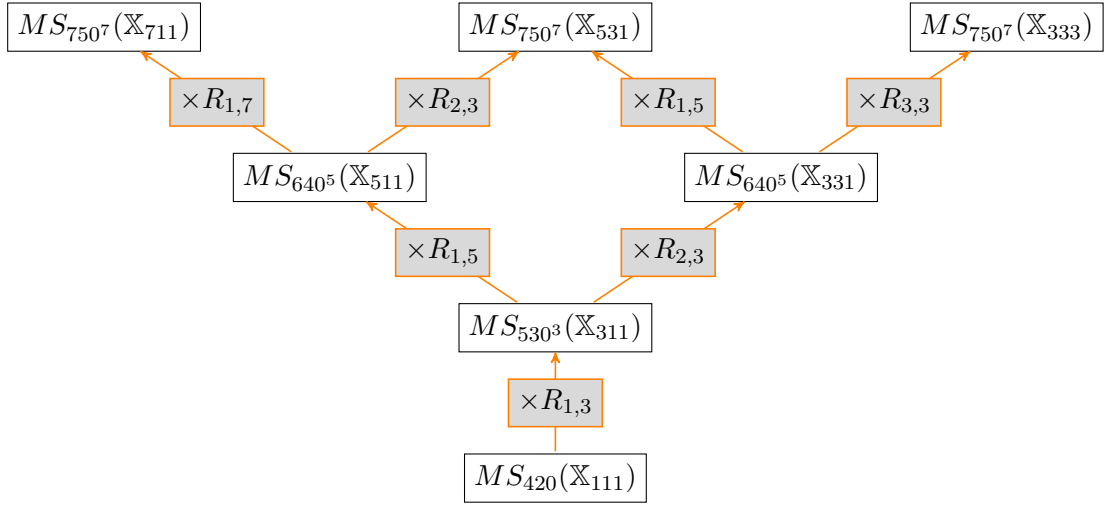

Let $\lambda=[\lambda_{1},\lambda_{2},\lambda_{3}]$
be a partition and $\mathbb X_{\lambda}={1-u^{\lambda_{1}}\over 1-u}x_{1}+{1-u^{\lambda_{2}}\over 1-u}x_{2}+{1-u^{\lambda_{3}}\over 1-u}x_{3}$.
We consider $\rho_{i}$ the operator adding $2$ to the $i$th entry of $\lambda$ if, after this operation, 
the resulting vector still is a partition.
We set also $R_{i;k}=\displaystyle\prod_{1\leq j\leq N\atop j\neq i}(u^{k}x_{i}-x_{j})$. Starting with $MS_{[420]}(\mathbb X_{111};q=u^{-2},t=u)$, and by 
$\mathbb X_{111}\rho_{1}=\mathbb X_{311}$, one obtains 
\[MS_{[530^{4}]}(\mathbb X_{311};q=u^{-2},t=u)=R_{1;3}MS_{[420]}(\mathbb X_{111};q=u^{-2},t=u).\]
The next step is more interesting because there are two kinds of specializations that provide nice factorizations:
\[{}
MS_{[640^{6}]}(\mathbb X_{511};q=u^{-2},t=u)\mathop=^{(*)}R_{1;5}MS_{[530^{4}]}(\mathbb X_{311};q=u^{-2},t=u),
\]
and
\[{}
MS_{[640^{6}]}(\mathbb X_{331};q=u^{-2},t=u)\mathop=^{(*)}R_{2;3}MS_{[530^{4}]}(\mathbb X_{311};q=u^{-2},t=u).
\]
{}
Continuing, one finds
\[{}
MS_{[750^{8}]}(\mathbb X_{711};q=u^{-2},t=u)\mathop=^{(*)}R_{1;7}MS_{[640^{4}]}(\mathbb X_{511};q=u^{-2},t=u),
\]
{}
\begin{multline*}
MS_{[750^{8}]}(\mathbb X_{531};q=u^{-2},t=u)\displaystyle\mathop=^{(*)}R_{2;3}MS_{[640^{4}]}(\mathbb X_{511};q=u^{-2},t=u)\displaystyle\mathop=^{(*)} \\
\displaystyle\mathop=^{(*)}R_{1;5}MS_{[640^{4}]}(\mathbb X_{331};q=u^{-2},t=u),
\end{multline*}
and
\[{}
MS_{[750^{8}]}(\mathbb X_{333};q=u^{-2},t=u)\mathop=^{(*)}R_{3;3}MS_{[640^{4}]}(\mathbb X_{331};q=u^{-2},t=u).
\]
The computation can be graphically represented as in Figure \ref{FigComp}.

Using vanishing properties it is not too difficult to show that when ${\lambda_{1}+\lambda_{2}+\lambda_{3}-3\over 2}=r$, then
$MS_{\left[4+r, 2+r, 0^{1+2r}\right]}(\mathbb X_{\lambda};q=u^{-2},t=u)$ factorizes nicely, and that if $\lambda\rho_{i}$ is well defined, one has
\begin{equation*}
	MS_{\left[4+(r+1), 2+(r+1), 0^{1+2r}\right]}(\mathbb X_{\lambda\rho_{i}};q^{-2},t)\mathop=^{(*)}R_{i,\lambda\rho_{i}[i]}MS_{\left[4+r, 2+r, 0^{1+2r}\right]}(\mathbb X_{\lambda};q=u^{-2},t=u).
\end{equation*}
This kind of formula takes place in a wider picture which will be investigated in a future paper.
\subsection{Factorizations of non symmetric Macdonald polynomials}
Specializations of singular non symmetric Macdonald polynomials   factorize as shown in the following examples:
\begin{equation*}
E_{210}(x_1,x_2,x_3;q=u^{-2},t=u)\mathop=^{(*)}(ux_2-x_1)(ux_3-x_1)(ux_3-x_2)\end{equation*}
\begin{equation*}
\begin{array}{r}\displaystyle
E_{630}(x_1,x_2,x_3;q=u^{-2},t=u^3)\mathop=^{(*)}
\left({\it x_2}\,u-{\it x_3} \right)  \left( -u{\it x_3}+{\it 
x_2} \right)  \left( {\it x_2}-{u}^{3}{\it x_3} \right)  \left( {\it x_1}
\,u-{\it x_3} \right)\\
 \left( -z{\it x_3}+{\it x_1} \right)  \left( {\it 
x_1}-{u}^{3}{\it x_3} \right)  \left( {\it x_1}\,u-{\it x_2} \right) 
 \left( {\it x_1}-{\it x_2}\,u \right)  \left( {\it x_1}-{\it x_2}\,{u}^{3
}\right),
\end{array}
\end{equation*}
\begin{equation*}
E_{420}(x_1,x_2,x_3;q=-t,t)\mathop=^{(*)}t(x_2 + x_3) (-t x_3 + x_2) (x_3 + x_1) (-t x_3 + x_1) (x_1 + x_2) (x_1 - x_2 t),
\end{equation*}
\begin{equation*}
\begin{array}{r}
E_{221100}(x_1,x_2,y_1,ty_1,y_2,ty_2;q=t^{-3},t)\displaystyle\mathop=^{(*)}(y_1 - y_2 t^2 ) (y_1 - y_2 t) (x_2 - y_2 t^2 )\\ (x_2 - t^2  y_1) (x_1 - y_2 t^2 ) (x_1 - t^2  y_1),\end{array}
\end{equation*}
\begin{equation*}
\begin{array}{r}
E_{42200}(x_1,y_1,y_1u^{2},y_2,y_2u^{2};q=u^{-3},t=u^{2})\displaystyle\mathop=^{(*)}
(y_1-y_2u^4)(y_1-y_2u)(y_1u-y_2)\\(y_1-y_2u^2)
(x_1-y_2u^4)(x_1-y_2u)(x_1-y_1u^4)(x_1-y_1u),
\end{array}
\end{equation*}
\begin{equation*}
E_{0022}(ty,y,x_3,x_4;q=t^{-3},t)\mathop=^{(*)}(tx_4-y)(tx_3-y)x_4x_3.
\end{equation*}
Notice that the last example is not singular and we have  $$E_{0022}(ty,y,x_3,x_4;q=t^{-3},t)\displaystyle\mathop{\neq}^{(*)}
M_{0022}(ty,y,x_3,x_4;q=t^{-3},t).$$ Indeed, $$M_{0022}(ty,y,x_3,x_4;q=t^{-3},t)\mathop=^{(*)}(tx_4-y)(tx_3-y)(x_4-1)(x_3-1).$$
But it is deduced from the singular polynomial
\begin{equation*}
E_{1100}(ty,y,x_3,x_4;q=t^{-3},t)\mathop=^{(*)}(tx_4-y)(tx_3-y),
\end{equation*}
by applying two times the affine operation. Some other examples involving vectors which are not partitions are more interesting.
 For instance, the polynomial $E_{1010}(x_{1},x_{2},x_{3},x_4;q=t^{-3},t)$ is singular and we have
\begin{equation*}
	E_{1010}(ty,x_{2},y,x_4;q=t^{-3},t)\mathop=^{(*)}(tx_4-y)(tx_2-y).
\end{equation*}
More general formulas for quasistaircases are also observed. For instance,
\begin{equation*}
E_{32000}(x_{1},x_{2},y,ty,t^{2}y;q=t^{-2},t)\mathop=^{(*)}
\left( {\it x_2}-y{t}^{3} \right)  \left( {\it x_2}-ty \right) 
 \left( {\it x_1}-y{t}^{3} \right)  \left( {\it x_1}-ty \right) 
 \left( {\it x_1}-t{\it x_{2}} \right).
\end{equation*}
\begin{equation*}\begin{array}{rcl}
	E_{4300000}(x_{1},x_{2},y,ty,t^{2}y,t^{3}y,t^{4}y;q=t^{-2},t)&\displaystyle\mathop=^{(*)}&
	\left( {\it x_{2}}-ty \right)  \left( {\it x_{2}}-y{t}^{3} \right) 
 \left( {\it x_{2}}-y{t}^{5} \right)  \left( {\it x_1}-y{t}^{3} \right)\\&&\times
 \left( {\it x_{1}}-y{t}^{5} \right)  \left( {\it x_{1}}-ty \right) 
 \left( {\it x_{1}}-t{\it x_{2}} \right). \end{array}
\end{equation*}
Numerical evidences suggest that one has a formula very close to those of Theorem \ref{LastFactTheo} but for a specialization under the form
\[{}
(t,q)=\left(u^{s\over g},u^{-{\ell+1\over g}}\omega_{1}\right)
\]
where $g=\gcd(\ell+1,s)$ and $\omega_{1}^{s\over g}$ is a $g$th primitive root of the unity.
Also as in section \ref{other}, we observe factorizations for other specializations of the variables $x_{i}$'s. For instance
\begin{equation*}\begin{array}{rcl}
	E_{4300000}(x_{1},t^{2}y_{1},y_{1},ty_{1},y_{2},ty_{2},t^{2}y_{2};q=t^{-2},t)&\displaystyle\mathop=^{(*)}&
	\left( {\it y_{1}}-t{\it y_2} \right)  \left( {\it y_{1}}-{t}^{3}{\it y_{2}}
 \right)  \left( t{\it y_{1}}-{\it y_{2}} \right)\\&&\times  \left( {\it x_{1}}-t{\it y_{2}
} \right)  \left( {\it x_{1}}-{t}^{3}{\it y_{2}} \right)  \left( {\it x_{1}}-
t{\it y_{1}} \right)  \left( {\it x_{1}}-{t}^{3}{\it y_{1}} \right) . \end{array}
\end{equation*}

The precise statements, the proofs, and the connection with the factorizations of symmetric Macdonald polynomials remain to be investigated.\\ \\
{\bf Acknowledgments:} The paper is partially supported by the GRR PROJECT MOUSTIC.
We thank Stephen Griffeth for pointing out the relevance of Theorem 1.1 in \cite{BGS} to this paper.
JGL gratefully acknowledges Thierry Jolicoeur for fruitful discussions.
 \bibliographystyle{plain}
 \bibliography{biblio_Fact}
\end{document}